\documentclass[twocolumn,showpacs,preprintnumbers,amsmath,prd,amssymb,superscriptaddress]{revtex4}

\usepackage{graphicx}
\usepackage{dcolumn}
\usepackage{bm}
\usepackage{multirow}
\usepackage{color}%

\def\nuebar{{\rm \bar{\nu}_e}}
\def\nue{{\rm \nu_e}}
\def\s2tw{{\rm sin ^2 \theta_{W}}}
\def\nuchrad{{\rm \langle r_{\bar{\nu}_e}^2\rangle}}
\def\munu{{\rm \mu_{\nu}}}

\begin{document}

\preprint{AS-TEXONO/09-04}

\title{
Measurement of $\nuebar$-Electron Scattering Cross-Section
with a CsI(Tl) Scintillating Crystal Array
at the Kuo-Sheng Nuclear Power Reactor
}

%

\newcommand{\as}{Institute of Physics, Academia Sinica, Taipei 11529, Taiwan.}
\newcommand{\ntu}{Department of Physics, National Taiwan University,
Taipei 106, Taiwan.}
\newcommand{\ihep}{Institute of High Energy Physics,
Chinese Academy of Science, Beijing 100039, China.}
\newcommand{\thu}{Department of Engineering Physics, Tsing Hua University,
Beijing 100084, China.}
\newcommand{\ks}{Kuo-Sheng Nuclear Power Station,
Taiwan Power Company, Kuo-Sheng 207, Taiwan.}
\newcommand{\iner}{Institute of Nuclear Energy Research,
Lung-Tan 325, Taiwan.}
\newcommand{\ciae}{Department of Nuclear Physics,
Institute of Atomic Energy, Beijing 102413, China.}
\newcommand{\metu}{Department of Physics,
Middle East Technical University, Ankara 06531, Turkey.}
\newcommand{\bhu}{Department of Physics,
Banaras Hindu University, Varanasi 221005, India.}
\newcommand{\natinstr}{National Instruments, 
Taipei 106, Taiwan.}
\newcommand{\nku}{Department of Physics, Nankai University,
Tianjin 300071, China.}
\newcommand{\umd}{Department of Physics, University of Maryland,
College Park MD 20742, U.S.A.}
\newcommand{\corr}{htwong@phys.sinica.edu.tw;
Tel:+886-2-2789-9682; FAX:+886-2-2788-9828.}

\author{ M.~Deniz }  \affiliation{ \as } \affiliation{ \metu }
\author{ S.T.~Lin } \affiliation{ \as }
\author{ V.~Singh }  \affiliation{ \as } \affiliation{ \bhu }
\author{ J.~Li }  \affiliation{ \as } \affiliation{ \ihep } \affiliation{ \thu }
\author{ H.T.~Wong } \altaffiliation[Corresponding Author: ]{ \corr } \affiliation{ \as }
\author{ S.~Bilmis }  \affiliation{ \as } \affiliation{ \metu }
\author{ C.Y.~Chang } \affiliation{ \as } \affiliation{ \umd }
\author{ H.M.~Chang } \affiliation{ \as }
\author{ W.C.~Chang } \affiliation{ \as }
\author{ C.P.~Chen } \affiliation{ \as }
\author{ M.H.~Chou } \affiliation{ \as }
\author{ K.J.~Dong } \affiliation{ \ciae }
\author{ J.M.~Fang } \affiliation{ \ks }
\author{ C.H.~Hu } \affiliation{ \iner }
\author{ G.C.~Jon } \affiliation{ \as }
\author{ W.S.~Kuo } \affiliation{ \iner }
\author{ W.P.~Lai } \affiliation{ \as }
\author{ F.S.~Lee } \affiliation{ \as }
\author{ S.C.~Lee } \affiliation{ \as }
\author{ H.B.~Li }  \affiliation{ \as }
\author{ H.Y~Liao } \affiliation{ \as }
\author{ C.W.~Lin } \affiliation{ \as }
\author{ F.K.~Lin } \affiliation{ \as }
\author{ S.K.~Lin } \affiliation{ \as }
\author{ Y.~Liu }  \affiliation{ \as } \affiliation{ \ihep }
\author{ J.F.~Qiu }  \affiliation{ \as } \affiliation{ \ihep }
\author{ M.~Serin } \affiliation{ \metu }
\author{ H.Y.~Sheng } \affiliation{ \as } \affiliation{ \ihep }
\author{ L.~Singh }  \affiliation{ \as } \affiliation{ \bhu }
\author{ R.F.~Su } \affiliation{ \ks }
\author{ W.S.~Tong } \affiliation{ \iner }
\author{ J.J.~Wang }  \affiliation{ \as }
\author{ P.L.~Wang } \affiliation{ \as } \affiliation{ \ihep }
\author{ S.C.~Wu } \affiliation{ \as } \affiliation{ \natinstr }
\author{ S.W.~Yang } \affiliation{ \as }
\author{ C.X.~Yu } \affiliation{ \nku }
\author{ Q.~Yue } \affiliation{ \thu }
\author{ M.~Zeyrek } \affiliation{ \metu }
\author{ D.X.~Zhao } \affiliation{ \as } \affiliation{ \ihep }
\author{ Z.Y.~Zhou } \affiliation{ \ciae }
\author{ Y.F.~Zhu }   \affiliation{ \as } \affiliation{ \thu }
\author{ B.A.~Zhuang } \affiliation{ \as } \affiliation{ \ihep }

\collaboration{TEXONO Collaboration}

\noaffiliation


\date{\today}

\begin{abstract}

The $\bar{\nu}_{e}-e^{-}$ elastic scattering 
cross-section was measured with
a CsI(Tl) scintillating crystal array 
having a total mass of 187~kg.
The detector was exposed to an average 
reactor $\bar{\nu}_{e}$ flux of 
$\rm{6.4\times 10^{12} ~ cm^{-2}s^{-1}}$
at the Kuo-Sheng Nuclear Power Station. 
The experimental design, conceptual merits,
detector hardware, data analysis and background
understanding of the experiment are presented. 
Using 29882/7369 kg-days of Reactor ON/OFF data, 
the Standard Model (SM) electroweak interaction 
was probed at the squared 4-momentum transfer range of
$\rm{Q^2 \sim 3 \times 10^{-6} ~ GeV^2}$.
The ratio of experimental to SM cross-sections of 
$ \xi =[ 1.08 \pm 0.21 (stat) \pm 0.16 (sys)] $ 
was measured. Constraints on the electroweak parameters
$( g_V , g_A )$ were placed, corresponding to
a weak mixing angle measurement of 
$ \s2tw = 0.251 \pm 0.031 ({\it stat}) \pm 0.024 ({\it sys})  $.
Destructive interference in the SM  
$\nuebar -$e process was verified.
Bounds on anomalous neutrino electromagnetic properties
were placed: neutrino magnetic moment at
$\mu_{\nuebar}< 2.2 \times 10^{-10} \mu_{\rm B}$ 
and the neutrino charge radius at 
$-2.1 \times 10^{-32} ~{\rm cm^{2}} 
< \nuchrad < 
3.3 \times 10^{-32} ~{\rm cm^{2}}$, 
both at $90 \% $ confidence level. 
\end{abstract}


\maketitle

\section{
Introduction
}

The compelling evidence of neutrino oscillations
from the solar, atmospheric as well as long baseline
accelerator and reactor neutrino measurements
implies finite neutrino masses and mixings~\cite{pdg08numix}.
Their physical origin and experimental consequences
are not fully understood.
Experimental studies on the neutrino properties
and interactions are crucial because
they can shed light to these 
fundamental questions and 
may provide hints or constraints to
models on new physics.

We report a study of neutrino-electron
scattering using reactor neutrinos 
at the Kuo-Sheng Nuclear Power Station
with a CsI(Tl) scintillating
crystal array.
The cross-section formulae are summarized in 
Section~\ref{sect::formulae}.
The conceptual design, hardware construction
and performance are presented in Section~\ref{sect::setup},
followed by discussions on event reconstruction,
background understanding and suppression, 
as well as experimental systematic effects.
Section~\ref{sect::results} shows 
results on the Standard Model (SM) 
electroweak physics~\cite{pdg08sm} as well as 
constraints on possible neutrino electromagnetic
interactions.

\section{Neutrino-Electron Scattering}
\label{sect::formulae}

Neutrino-electron scattering has been studied 
with several generations of experiments at 
the accelerator using mostly muon-neutrinos 
$\nu_{\mu} ( \bar{\nu}_{\mu} )$~\cite{nuecrosssection,numuele}.
It is a pure leptonic process and therefore provides a clean
test to SM.
The typical squared 4-momentum transfer was
$Q^2 \sim 10^{-2}~ {\rm GeV^2}$ 
and the electroweak angle $\s2tw$ was probed to an accuracy
of $\pm$3.6\%.

\begin{figure}
\begin{center}
\includegraphics[width=8cm]{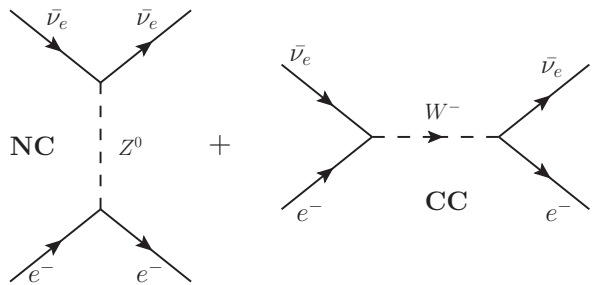}
\caption{\label{feydiag}
Interactions of $\nuebar$ with electron via the
SM-allowed charged current (CC)  and 
neutral current (NC) channels. 
There is in addition interference effect between them.
}
\end{center} 
\end{figure}

\begin{table*}
\caption{
Summary of published
$\nue -$ and $\nuebar - e$
scattering
cross-section and $\s2tw$ measurements.
Unavailable entries are denoted by ``N/A''.
}
\label{summarynueexpt}
\begin{ruledtabular}
\begin{tabular}{lccccc}
Experiment & $\rm{E_{\nu}}$ (MeV) &  T (MeV) & Events~\cite{rashba}
& Published Cross-Section & $\s2tw$ \\ \hline
\multicolumn{6}{l}{\underline{Accelerator $\nue$} : } \\
~~LAMPF~\cite{lampf} & $7 < {\rm E_{\nu}} < 50$ & 7$-$50  & 236 &
$[10.0\pm1.5\pm0.9] \cdot {\rm E}_{\nu}$ & 
$0.249\pm0.063$ \\
& & & &  $\times 10^{-45} ~ {\rm cm}^2$ & \\
~~LSND~\cite{lsnd} & $20 < {\rm E_{\nu}} < 50$ & 20$-$50  & 191 &
$[10.1\pm1.1\pm1.0] \cdot {\rm E}_{\nu}$ & 
$0.248\pm0.051$ \\
& & & & $\times 10^{-45} ~ {\rm cm}^2$ & \\
\multicolumn{6}{l}{\underline{Reactor $\nuebar$} : } \\

~~Savannah River & & & & \\
& $1.5 < {\rm E_{\nu}} < 8.0$ & 1.5$-$3.0 & 381 & $[0.87\pm 0.25]
\cdot \sigma_{V-A}$ &  \multirow{2}*{ \} ~ 0.29$\pm$0.05 }\\
~~~\multirow{2}*{Original~\cite{savannah} ~~~~~~~ \{ } & $3.0 < {\rm
E_{\nu}} < 8.0$  & 3.0$-$4.5 & 77 &
$[1.70\pm 0.44] \cdot \sigma_{V-A}$ & \\

& $1.5 < {\rm E_{\nu}} < 8.0$ & 1.5$-$3.0 & N/A &
$[1.35\pm 0.4]\ \cdot \sigma_{\rm SM}$ & \multirow{2}*{\} ~  N/A } \\
~~~\multirow{2}*{Re-analysis~\cite{vogelengel} ~~~ \{ } & $3.0 <
{\rm E_{\nu}} < 8.0$  & 3.0$-$4.5 & N/A &
$[2.0\pm 0.5] \cdot \sigma_{\rm SM}$ &  \\

~~Krasnoyarsk~\cite{kras} &
 $3.2 < {\rm E_{\nu}} < 8.0$  & 3.2$-$5.2 & N/A &
$[4.5\pm 2.4]$ & $0.22_{-0.8}^{+0.7}$ \\
& & & &  $\times 10^{-46} ~ {\rm cm}^2 / {\rm fission}$ & \\
~~Rovno~\cite{rovno} &
 $0.6 < {\rm E_{\nu}} < 8.0$ & 0.6$-$2.0 & 41 &
$[1.26\pm 0.62]$ & N/A \\
& & & &  $\times 10^{-44} ~ {\rm cm}^2 / {\rm fission}$ & \\
~~MUNU~\cite{munu} &
 $0.7 < {\rm E_{\nu}} < 8.0$  & 0.7$-$2.0 & 68 &
$ [ 1.07 \pm 0.34 ] $ events/day & N/A \\
~~TEXONO (This Work) & $3.0 < {\rm E_{\nu}} < 8.0$ & 3.0$-$8.0 
& 414$\pm$80$\pm$61
& $[1.08\pm0.21\pm0.16] \cdot \sigma_{\rm SM}$ 
& $0.251 \pm 0.031 \pm 0.024 $
\end{tabular}
\end{ruledtabular}
\end{table*}

Using electron-neutrinos as probe, the interaction
\begin{equation}
\rm{
\nue (\nuebar) + e^- 
\rightarrow 
\nue (\nuebar) + e^- 
}
\end{equation}
has been studied at medium energy accelerators~\cite{lampf,lsnd}
as well as at the power reactors~\cite{savannah,kras,rovno,munu}.
It is also an important channel in the detection
of solar neutrinos~\cite{snu} where the 
SM $\nue -$e scattering cross-section was
used to extract neutrino oscillation parameters.
This process is among the few of the SM interactions
which proceed via charged current (CC), neutral current (NC)
as well as their interference (Int)~\cite{kayser79}, 
as illustrated schematically in Figure~\ref{feydiag}.
The interference effect in
$\nue -$e scattering is the origin of 
matter oscillation of solar neutrinos in the
interior of the Sun~\cite{pdg08numix}.

The experimental results on
$\nue -$ and $\nuebar - e$ scattering
are summarized in Table~\ref{summarynueexpt}.
Neutrino-electron scattering was first observed
with reactor reactors in the Savannah River
experiment~\cite{savannah}.
Re-analysis of the data 
by a later work~\cite{vogelengel} with improved input
on the reactor neutrino spectra and electroweak parameters
gave cross-sections
which were about $2 \sigma$ higher than the SM values. 
The discrepancies were interpreted as hints of
anomalous neutrino interactions.
Other subsequent experiments~\cite{kras,rovno,munu}
focused on the searches of 
neutrino magnetic moments at low recoil energy
such that their sensitivities to SM physics
were limited.

\subsection{Electroweak Parameters}

The SM
differential cross-section in the
laboratory frame for 
$\nu_{\mu} ( \bar{\nu}_{\mu} ) -$e
elastic scattering, where only NC is involved,
is given by~\cite{pdg08sm,nuecrosssection};
\begin{eqnarray}
\left[ \frac{d\sigma}{dT} ( ^{[-]} \hspace*{-0.35cm} {\nu}_{\mu} e) 
\right] _{SM} 
& = &  \frac{G_{F}^{2}m_{e}}{2\pi }  \cdot 
[ ~ \left(g_{V} \pm g_{A} \right) ^{2}  \nonumber \\
& + &  \left( g_{V} \mp g_{A} \right) ^{2}\left(1-
\frac{T}{E_{\nu }}\right) ^{2}  \nonumber  \\
& - & ( g_{V}^2 - g_{A}^2 ) ~ \frac{m_{e}T}{E_{\nu}^{2}} ~ ] ~~~ , 
\label{eq_cs_numu}
\end{eqnarray}
where $G_F$ is the Fermi coupling constant,
$T$ is the kinetic energy of the recoil electron, 
$E_{\nu }$ is the incident neutrino energy 
and $g_{V}$, $g_{A}$ are, respectively, the
vector and axial-vector coupling constants.
The upper(lower) sign refers to the
interactions with $\nu_{\mu} ( \bar{\nu}_{\mu} )$.
For $\nu_{e} ( \nuebar ) -$e scattering,
all CC, NC and Int are involved~\cite{kayser79}, 
and the cross-section
can be obtained by making the replacement 
$g_{V,A} \rightarrow ( g_{V,A} + 1 )$. 
In the case of $\nuebar -$e which is 
relevant for reactor neutrinos,
\begin{eqnarray}
\left[ \frac{d\sigma}{dT}(\bar{\nu}_{e}e ) \right] _{SM} & = & 
\frac{G_{F}^{2}m_{e}}{2\pi }  \cdot  
[ ~ \left(g_{V}-g_{A}\right) ^{2}  \nonumber \\
& + & \left( g_{V}+g_{A}+2\right) ^{2}\left(1-
\frac{T}{E_{\nu }}\right) ^{2}  \nonumber  \\
& - & (g_{V}-g_{A})(g_{V}+g_{A}
+2)\frac{m_{e}T}
{E_{\nu}^{2}}  ~ ] .
\label{eq::gvga}
\end{eqnarray}
The SM assignments to the coupling constants
are:
\begin{equation}
g_{V}=-\frac{1}{2}+2\sin ^{2}\theta _{W}\text{ \ \ \ \ and \ \ \ \ }
g_{A}=-\frac{1}{2}\label{eq_gvga} ~~~ ,
\end{equation}
where $\s2tw$ is the weak mixing angle.
The SM differential cross-section expressed
in terms of $\s2tw$ is accordingly:
\begin{eqnarray}
\left[ \frac{d\sigma}{dT}(\bar{\nu}_{e}e ) \right] _{SM} & = & 
\frac{G_{F}^{2}m_{e}}{2\pi } ~ \cdot \nonumber \\
& \{  &
4 ~ ( \s2tw ) ^2  \left[  1+ \left(
1-\frac{T}{E_{\nu }}\right) ^{2} -
\frac{m_{e}T}{E_{\nu }^{2}} \right] \nonumber \\
& + & 4 ~ \sin ^{2}\theta _{W}  \left[  \left( 1-\frac{T}{E_{\nu }}\right) ^{2}
- \frac{m_{e}T}{2E_{\nu }^{2}} \right]  \nonumber \\
& + & \left( 1-\frac{T}{E_{\nu }}\right) ^{2} ~ \}  ~~ .
\label{eq::s2tw}
\end{eqnarray}

The observables in an experiment
are the event rates ($R_{expt}$).
The SM predicted rate, 
expressed in unit of ${\rm kg^{-1} day^{-1}}$,
can be written as
\begin{equation}
R_{SM} ( \nu ) ~ = ~ \rho_e ~
\int_{T} \int_{E_{\nu }} 
[ \frac{d\sigma}{dT} ] _{SM} ~
\frac{d\phi}{dE_{\nu}} ~ dE_{\nu} ~ dT  ~~ ,
\label{eq::sm}
\end{equation}
where $\rho_e$ is the electron number density 
per kg of target mass, 
and $d \phi _{\nu} / d E_{\nu}$ denotes the
neutrino spectrum.

Results of this work are reported in several schemes
using $R_{expt}$.
Firstly, the cross-section ratio 
\begin{equation}
\xi = \frac{R_{expt} ( \nu )}{R_{SM} ( \nu )} 
\label{eq::xidef}
\end{equation}
can be used to probe 
new physics in a model-independent way.
Alternatively, taking SM electroweak interactions 
but allowing the parameters to assume any values,  
the allowed ranges of
$( g_V , g_A )$ 
as well as 
$\s2tw$ 
can be derived
from $R_{expt}$, following
Eqs.~\ref{eq::gvga}\&\ref{eq::s2tw}, respectively.

To study the interference effects,
the measured rate can be expressed as
\begin{equation}
\label{eq::interf}
R_{expt} = R_{CC} + R_{NC} + \eta \cdot R_{Int}  ~~ .
\end{equation}
The CC$-$NC interference for $\nue(\nuebar) -$e 
is destructive in SM, or equivalently
$\eta (SM)  = -1 $. 
Possible deviations in
the sign and magnitude of 
the interference effects ($\eta$)
can be probed.

It follows from Eqs.~\ref{eq::s2tw}\&\ref{eq::sm}
and the analogous formulae for $\nue -$e that,
under realistic experimental configurations,
the projected accuracies on $\s2tw$ 
(denoted by $\Delta [ \s2tw ]$) are related to
the experimental uncertainties in $\xi$ 
(denoted by $\Delta [ \xi ]$) 
by:
\begin{equation}
\label{eq::ds2tw}
\Delta [ \s2tw ] ~ \sim ~ 
\{ ~
\begin{array}{lr}
0.15 \cdot ~ \Delta   [ \xi ( \nuebar e ) ] & \\ 
0.35 \cdot  ~ \Delta   [ \xi ( \nue e ) ]  & 
\end{array}
\end{equation}
for reactor $\nuebar -$e (this work) and
accelerator $\nue -$e~\cite{lampf,lsnd} experiments, 
respectively. 
Accordingly, the studies of reactor $\nuebar -$e 
are expected to improve on 
the sensitivities of $\s2tw$
and $( g_V , g_A )$
at the same experimental accuracies
as those from $\nue -$e measurements.
The relative strength of the three components
normalized to $R_{expt} = 1$
are in the ratios of
\begin{equation}
\label{eq::csratio}
\begin{array}{lll}
\multirow{2}*{  $( R_{CC} : R_{NC} : R_{Int} ) \sim $ \{} 
& ( 0.77 : 0.92 : 0.69 )  & 
{\rm ~ for ~ } \nuebar - {\rm e} 
\\ 
& ( 1.77 : 0.16 : 0.93 ) &
{\rm ~ for ~ } \nue - {\rm e}  ~ . 
\end{array}
\end{equation}
The stronger NC component in
$\nuebar -$e scattering
is the physical basis of the sensitivity 
enhancement in the derivation of $\s2tw$.

The SM was tested and $\s2tw$
was precisely measured in the
high energy (${\rm Q^2 > GeV^2}$)
region with accelerator experiments on
$e^+e^-$, polarized $ep$ and $\nu N$ deep inelastic
processes, and in the
low energy  (${\rm Q^2 < 10^{-6} ~ GeV^2}$)
region with measurements on
atomic parity violation~\cite{pdg08sm}.
Among them, the $\s2tw$ derived from
the NuTeV experiment on $\nu N$
deep inelastic scattering~\cite{nutev}
was 3$\sigma$ higher than  SM prediction,
though the interpretations were complicated
by strong interaction effects~\cite{pdg08sm}.
Destructive interference
according to SM prediction
has been demonstrated
by accelerator $\nue -$e scattering
experiments~\cite{lampf,lsnd}.

The objective of this work is to
bridge the ${\rm Q^2}$ gap in probing SM electroweak physics
with reactor $\nuebar -$e interactions.
In particular,
the interference effects
are studied in this unique system.
This would complement
the precision data obtained at accelerator
at higher $Q^2$.
The measurements would place constraints on various 
anomalous neutrino interactions such as those
discussed in the next section.

\subsection{Neutrino Electromagnetic Properties}

The neutrino electromagnetic interactions~\cite{nuem} 
provide natural extensions to SM.
The relevant parameters are $\nuchrad$~\cite{rashba},
usually called the ``neutrino charge radius'',
and neutrino magnetic moments ($\munu$)~\cite{munureview} which
describe possible neutrino interactions with matter via the
exchange of virtual photons without and with
the change of its helicity, respectively.

Interpretations of $\nuchrad$ remain controversial. 
A straight-forward definition has been shown to
be gauge-dependent and hence $\nuchrad$ is unphysical~\cite{shrock}.
However, there are recent attempts to define
a physical observable with $\nuchrad$~\cite{bernabeu},
which give a predicted value of 
$\rm{\langle r_{\nu_e}^2\rangle = 0.4 \times 10^{-32} ~ cm^2}$
with the SM framework.
We adopt in this article the more general interpretation that
$\nuchrad$ parametrizes contributions to non-standard interactions
in neutrino scattering~\cite{qradpdg08}.

Changes to the SM cross-sections due to $\nuchrad$
can be obtained from Eq.~\ref{eq::s2tw}
via the replacement~\cite{vogelengel}:
\begin{equation}
\s2tw  \rightarrow \s2tw + 
(\frac{\sqrt{2} \pi \alpha_{em}}{3 G_F}) \nuchrad ~~ ,
\label{eq::qrad}
\end{equation}
where $\alpha_{em}$ is
the fine structure constant. 

Contributions of $\munu$ can be described by an additional
term to Eqs.~\ref{eq::gvga}\&\ref{eq::s2tw}:
\begin{equation}
\left( \frac{d\sigma }{dT}\right) _{\mu _{\nu }}=\frac{\pi \alpha
_{em}^{2}\mu _{\nu }^{2}}{m_{e}^{2}}
\left[ \frac{1-T/E_{\nu }}{T}\right] ~~ .
\label{eq_mm}
\end{equation}
The SM prediction of $\munu$  
for massive Dirac neutrinos is extremely
small ($3.2 \times 10^{-19} ~ \mu_{\rm B}$ 
where $\mu_{\rm B}$ is the Bohr magneton).
However, various models with Majorana neutrinos 
can give rise to $\munu$ at the 
range of $ ( 10^{-10} - 10^{-13} ) ~ \mu_{\rm B}$
relevant to experiments and astrophysics~\cite{munureview}.
The most sensitive direct laboratory limits on $\munu$
come from high-purity germanium detectors 
at about 10~keV threshold
with reactor $\nuebar$~\cite{texonomunu,gemma}.
At this low recoil energy,
the $\munu$ contributions at the present limit
are orders of magnitude
larger than those due to SM $\nuebar -$e cross-sections.

\section{Experimental Set-Up}
\label{sect::setup}

\subsection{Laboratory Facilities and Neutrino Flux}

\begin{table*}
\caption{
Summary of the key information of the four data taking periods. 
The period numbering follows the same scheme as in Ref~\cite{texonomunu} .}
\label{tab_periods}
\begin{ruledtabular}
\begin{tabular}{lcccccccc}
& Data Taking  & Reactor ON & {Reactor OFF} &
DAQ & DAQ & Average $\nuebar$ & Fiducial  \\
Period & Calender Time &  Live Time &  Live Time
& Live Time & Threshold & Flux & Mass \\
& & (days) & (days) & $(\%)$ &  (keV) & $(10^{12} cm^{-2}s^{-1})$ &
(kg) \\ \hline
II & Feb. 2003 - Oct. 2003 &  95.2 &  48.4 & 88.8 &
100 & 6.27 & 43.5 \\
III & Sept. 2004 - Oct. 2005 &  192 &  36.6 & 93.4 &
500 & 6.50 & 40.5 \\
IV & Mar. 2006 - May 2007 &  204.9 &  43.5 & 88.0 &
500 & 6.44 & 51 \\
V & June 2007 - Feb. 2008 &  132.8 &  27.6 & 91.9 & 500 
& 6.29 & 57 \\ \hline
Combined & Feb. 2003 - Feb. 2008 &  624.9 &  156.1 &
90.4 & $-$ & 6.39 & $-$
\end{tabular}
\end{ruledtabular}
\end{table*}

\begin{figure}
\begin{center}
\includegraphics[width=8cm]{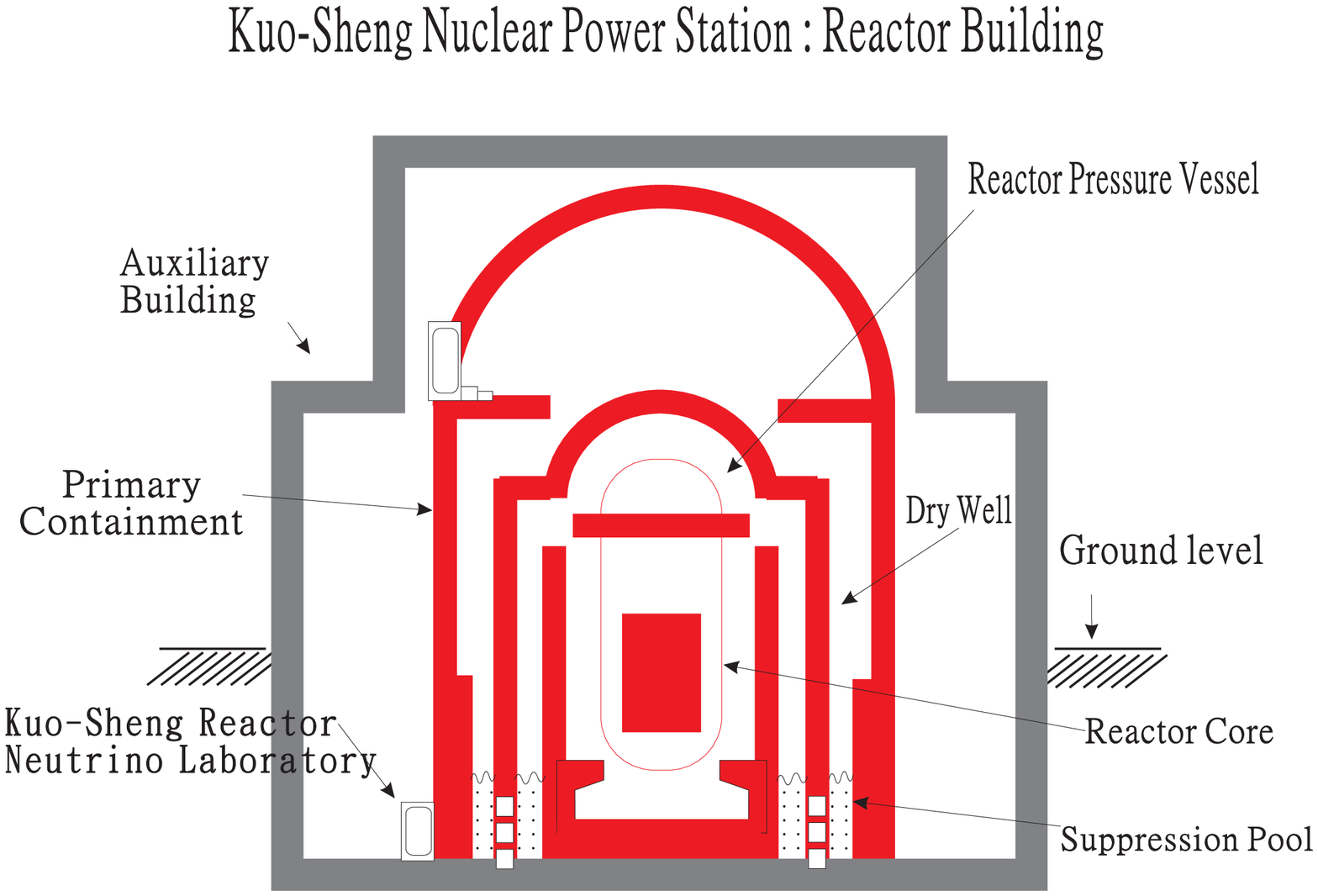}
\caption{\label{fig_reactor} 
Schematic layout of the Kuo-Sheng
Neutrino Laboratory together with the
reactor core and building.
}
\end{center}
\end{figure}

A research program
on low energy neutrino physics~\cite{ksprogram}
is being pursued by the TEXONO
Collaboration at the Kuo-Sheng Neutrino Laboratory (KSNL),
which is located at a distance of 28~m from
Core \#1 of the Kuo-Sheng Nuclear Power
Station in Taiwan.
A schematic diagram is depicted in Figure~\ref{fig_reactor}.
The site is at the ground floor of the reactor 
building at a depth of 10~m below ground level,
with an overburden of about 30~meter-water-equivalence.
The nominal thermal power output is 2.9~GW.
The standard operation includes
about 18~months of Reactor ON periods 
separated by 50~days of
Reactor outage OFF periods when typically
one-third of the fuel elements are replaced.

\begin{figure}[hbt]
\includegraphics[width=8.0cm]{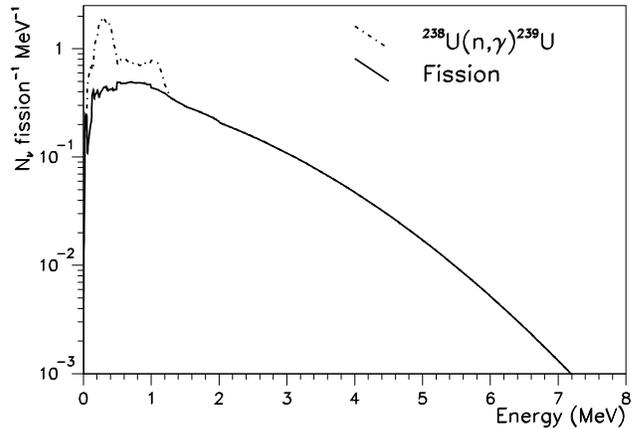}
\caption{
Total $\nuebar$ spectrum at typical power
reactor operation.
}
\label{rnusp}
\end{figure}

A summary of the key information on the four data
taking periods reported in this article is given
in Table~\ref{tab_periods}.
The evaluation of the reactor neutrino flux and spectra 
was discussed in details in Refs.~\cite{texonomunu,rnuloe}.
The average $\nuebar$-flux at KSNL is 
$6.4 \times 10^{12} ~ {\rm cm^{-2}s^{-1}}$.
A typical spectrum is displayed in Figure~\ref{rnusp}.
It has been demonstrated 
through $\nuebar -$proton measurements
that the integrated $\nuebar$-flux
for  $E_{\nu} > 1.8 ~ {\rm MeV}$~\cite{nuebarflux}
and $\nuebar$-spectra 
for $E_{\nu} > 3 ~ {\rm MeV}$~\cite{rnuhie}
agreed with calculations to better than
$<$3\%  and $<$5\%, respectively. 

\begin{figure}
\begin{center}
\includegraphics[width=8cm]{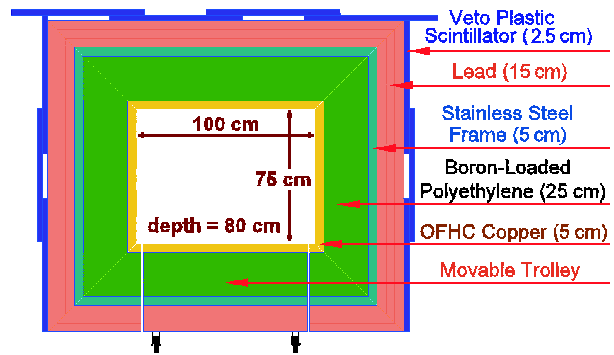}
\caption{\label{shielding} 
The shielding design of KSNL.
Similar structures apply to the back and
front walls.
Detectors and inner shieldings were
placed in the inner target volume.
}
\end{center}
\end{figure}

The laboratory is equipped
with a 50-ton shielding structure
depicted schematically in Figure~\ref{shielding},
consisting of, from outside in,
2.5~cm thick plastic scintillator panels with
photo-multiplier tubes (PMTs) readout
for cosmic-ray veto,
15~cm of lead, 5~cm of stainless
steel support structures, 25~cm of boron-loaded polyethylene
and 5~cm of Oxygen Free High Conductivity (OFHC) 
copper.
The inner target volume with a dimension of
100$\times$80$\times$75~$\rm{cm^3}$
allows different detectors for various physics topics
to be placed.
Data were taken
with a CsI(Tl) scintillating crystal array 
during data acquisition (DAQ) periods II$-$V.
Each period consisted of both reactor ON and OFF
data taking.

\subsection{Conceptual Design and Motivations}

The merits of scintillating crystal detectors
in generic low background low energy experiments have
been discussed~\cite{crystalmerits}.
This experiment adopted CsI(Tl) crystal scintillator
packed in a compact array as both target and detector.
A schematic layout is given in Figure~\ref{fig_csi}.
Several detector characteristics and
design features were incorporated~\cite{csinim01} 
which contributed
to the improvement in the experimental sensitivities:

\begin{figure}
\begin{center}
\includegraphics[width=8cm]{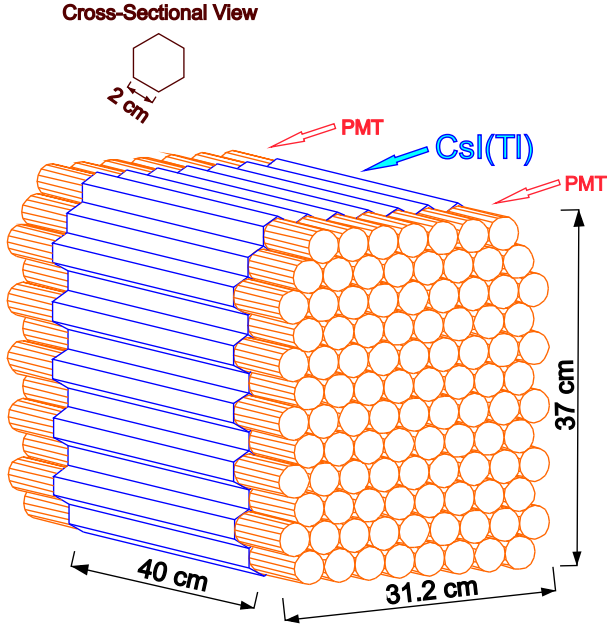}
\caption{\label{fig_csi} 
Schematic drawing of the CsI(Tl)
scintillating crystal array. 
Light output is recorded
by PMTs at both ends.
}
\end{center}
\end{figure}

\begin{description}
\item [(i) Proton-Free Target Region:]
The CsI(Tl) crystal is only weakly hygroscopic 
and does not require a hermetic container to seal it
from ambient humidity (in contrast to NaI(Tl) crystal).
The crystal is also mechanically stable and self-supporting.
Therefore, the target region was made up almost
entirely of CsI(Tl) 
(equal amount of Cs and I, with 
0.15\% admixture of Tl).
The other materials were the small amount 
of teflon wrapping sheets, made up of C and
F and contributing to only about 0.13\% by mass.
There were no protons, such that possible
neutrino-induced background from $\nuebar - $p
was eliminated. 
The cross-section of this interaction 
is $>$$10^2$ times higher 
than that of $\nuebar-$e. 
This background
could not be suppressed with Reactor ON/OFF 
comparisons, and could be a potential problem with the 
Savannah River experiment~\cite{savannah} where 
plastic scintillators were adopted as target.
\item [(ii) Completely Active Fiducial Volume:]
The absence of detector housing allowed a fiducial volume
which was totally active. The probability of
background events to be completely measured was enhanced,
and this was beneficial to background understanding
and suppression.
\item [(iii) Complete Three-Dimensional Reconstruction:]
Each CsI(Tl) crystal module consisted of 
a hexagonal-shaped cross-section with
2~cm side and a length of 40~cm, giving a
modular mass of 1.87~kg.
Scintillation photons were read out by PMTs
at both ends. The sum and difference 
of the two signals provided the energy
and position information, respectively.
A three-dimensional reconstruction of the 
events was achieved. These information greatly
enhanced the capabilities of background diagnostics 
and evaluation.
In particular, background induced by ambient radiations
was suppressed by rejecting events 
at the outer modules or close to the PMTs.  
The high atomic number for Cs and I (Z=51 and 53, respectively)
allowed efficient attenuation and therefore
compact detector geometry.
\item [(iv) Large Mass and Expandable Detector:]
This experiment was based on a
modular CsI(Tl) crystal array with a total mass
of 187~kg.
Such detector approach with similar target mass scale
was also adopted in cold dark matter searches
in the KIMS experiment~\cite{kims}.
The design can be easily
expanded to ton-scale experiments and beyond.
\item [(v) Pulse Shape Discrimination:]
The light emission profiles of CsI(Tl) offered
excellent pulse shape discrimination (PSD) between
$\gamma$/e events from those due to $\alpha$-particles
and nuclear recoils~\cite{csiproto,lepsd}.
This allowed precise measurements
of the internal contaminations for
background suppression and diagnostics.
\item [(vi) Focus at High Energy Events:]
The reactor $\nuebar$-spectra below
2~MeV has large uncertainties~\cite{rnuloe}, while ambient
background dominate below the natural radioactivity
end-point of 2.6~MeV.
Accordingly, only events
with $T > 3 ~ {\rm MeV}$ were studied as
potential $\nuebar -$e candidates. 
The low energy events were still recorded and analyzed
for the purposes of calibrations and background diagnostics.
\end{description} 

\subsection{Detector Construction and Readout} 

As depicted in Figure~\ref{fig_csi},
the scintillating CsI(Tl) crystal detector modules
were packed into a matrix array, with minimal
inactive dead space due to the teflon wrapping sheets.
The configurations varied between the different DAQ periods,
but the operation conditions were kept uniform and stable
within one period.
Therefore, each DAQ period can be taken as an independent
experiment.
At the end of data taking, a $12 \times 9$ array was deployed
giving a total mass of 187~kg.
Fiducial volume was defined to be the inner crystals
with a separation of $>$4~cm from the PMTs at both ends.
The fiducial masses 
for individual periods are given in Table~\ref{tab_periods}.

There were two types of crystal modules~\cite{csiproto} from
two production batches:
(a) single crystal with 40~cm length were used as target 
placed in the central region, while 
(b) two pieces of 20 cm long crystals optically glued together
were placed in the outer layers as active veto.
The light output was read out at both ends 
of the crystal modules 
by custom designed PMTs 
with low-activity glass and diameter of 29~mm.
The target array was housed inside a 
OFHC copper box of thickness 2.5~mm. Additional
copper shielding blocks were placed on top of the box 
to fill up the inner target volume of Figure~\ref{shielding}.
The box was flushed with dry nitrogen to purge
the radioactive radon gas. 
The CsI(Tl) array shared the target volume and
the downstream DAQ systems 
with germanium detectors for 
magnetic moment studies~\cite{texonomunu} and
dark matter searches~\cite{texonocdm}.

The electronics and DAQ systems 
were described in Ref.~\cite{ksdaq}.
The DAQ system was VME-based running on LINUX operating system.
The PMT signals were fed
to custom-built shaping amplifiers whose output
were recorded by 
Flash Analog-to-Digital Converter (FADC) modules 
at a clock rate of 20~MHz and 8-bit dynamic range. 
The DAQ trigger was generated by discriminator 
set at threshold
of 100~keV for P-II and 
500~keV for P-III,IV,V,
much lower than the relevant signal region.
Signals from all sub-dominant channels with
energy depositions $\agt 10 ~ {\rm keV}$,
as well as the PMT signals from the veto-panel system
and various control parameters, 
were also recorded.

A special feature of the DAQ system was the recording of any
events delayed as much as 500~$\mu$s after the initial trigger.
The delay time was measured with 1~$\mu$s resolution.
This allowed measurement of delayed-coincidence  
events due to internal radioactivity, which in turn was 
crucial to background diagnostics
and suppression.
The DAQ output was zero-suppressed, 
such that only those CsI(Tl) channels having signals
within (-5~$\mu$s, 500~$\mu$s) relative to the
trigger instant were recorded.

The DAQ dead time was accurately measured by the 
random trigger (RT) events generated at 0.1~Hz uncorrelated
with the rest of the hardware.
The typical trigger rate for the CsI(Tl) array was
20$-$30~Hz, corresponding 
to 8$-$12\% of DAQ dead time.  
Data were taken with the germanium system in parallel
in Periods~II$-$IV, but the additional contributions 
to the DAQ dead time were only minor.

\subsection{Detector Performance}

\begin{figure}
\begin{center}
\includegraphics[width=8cm]{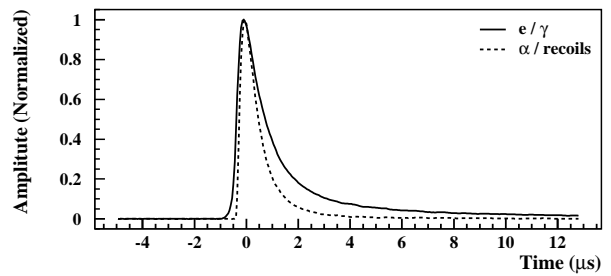}
\caption{\label{fig_ref_pulse}
Averaged pulse shapes
due to  5.4~MeV $\alpha$-particles
and $\gamma$-rays of 662~keV.
Nuclear recoils, as measured with scattering
with a neutron beam, give rise the same pulse
shapes as the $\alpha$-events~\cite{csiproto}.
}
\end{center}
\end{figure}

The intrinsic performance of the CsI(Tl) crystal modules
were discussed in details in Ref.~\cite{csiproto}.
The energy and position resolutions on individual
module measured with a $^{137}$Cs $\gamma$-source 
at 662~keV were
4\% and $<$1~cm root-mean-square (RMS), 
respectively.
The averaged pulse shapes
for $\gamma$/e  events and $\alpha$-particles
in CsI(Tl) are displayed in Figure~\ref{fig_ref_pulse}. 
Separation of $\gamma$/$\alpha$ at $>$99\% 
was achieved by PSD
down to 100~keV electron-equivalence energy.

The FADC has a hardware dynamic range of only 8-bit.
Software algorithm was devised to correct the saturated pulse
shapes~\cite{satpulse}. The effective range
was extended by 4 more bits without affecting the
performance parameters like energy resolution and PSD.
The CsI(Tl) output for the current measurements
typically saturated at about 2~MeV,
so that the events with energy $<$10~MeV relevant to the
analysis were well-reconstructed.

\section{Data Analysis}

\subsection{Light Collection}

\begin{figure}
\begin{center}
\includegraphics[width=8cm]{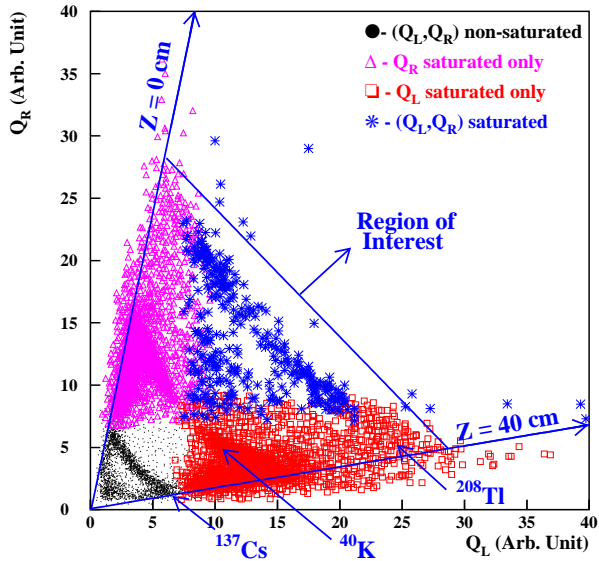}
\caption{
\label{fig_qlvsqr} 
Typical $Q_{L}$ versus $Q_{R}$
distribution for H1(CRV) events showing the background events
of natural sources. Different colors denote
whether the PMT signals are saturated at
their FADC readout or not.
Additional software routines were devised to 
provide correct energy information for saturated events.
}
\end{center}
\end{figure}

The raw input to subsequent analysis 
were the light output (denoted by $Q_L$ and $Q_R$)
derived by summing the pedestal-subtracted
FADC signals from the PMTs on both ends 
of the CsI(Tl) modules.
Depicted in Figure~\ref{fig_qlvsqr} is
a typical normalized $Q_L$ versus $Q_R$ distribution.
The selected events were those  
having signals only in one crystal (H1),
with ``cosmic-ray veto'' (CRV) imposed and
software correction applied to 
the saturated pulses~\cite{satpulse}.
The different color schemes denote the
status on pulse saturation of the two PMTs.

Three bands along the increasing energy axis
are conspicuous, corresponding to background
due to $\gamma$-rays 
from $^{137}$Cs (662~keV),
$^{40}$K (1461~keV) and $^{208}$Tl (2614~keV).
These lines were important for
{\it in situ} calibration as well
as background diagnostics.
The sharp reduction
of background beyond the $^{208}$Tl energy $-$ the
signal region of this measurement
$-$ is very distinct.
The enhanced event rates at both edges indicate that
most background sources were external 
to the detector.

\subsection{Event Reconstruction}

The objectives of event reconstruction were
to provide measurements on energy (E) and longitudinal position (Z)
using $Q_L$ and $Q_R$.
The calibration procedures were performed with
{\it in situ} data, typically once every week.

\begin{figure}
\begin{center}
\includegraphics[width=8cm]{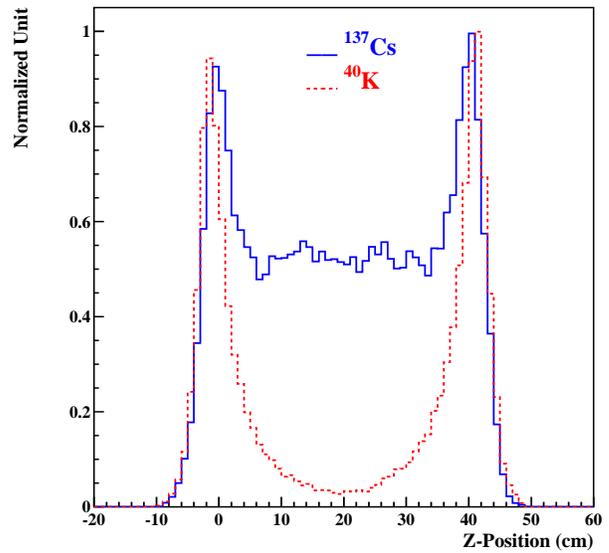}
\caption{
\label{zpos} 
The longitudinal Z-position distributions for
events at energy corresponding to
$\gamma$-lines of
$^{137}$Cs (solid histogram) and $^{40}$K (dotted histogram).
}
\end{center}
\end{figure}

The longitudinal Z-position for the $i^{th}$ crystal module is given by:
\begin{equation}
{\rm Z} ~ \propto ~ 
\left[
\frac{\beta_{i} \cdot Q_{R}-Q_{L}}{\beta_{i} \cdot Q_{R}+Q_{L}}  
\right] ~~ ,
\label{eq_zpos}
\end{equation}
where $\beta_{i}$'s are parameters
to absorb the residual difference in response between the
left and right readout.
The values of $\beta_i$'s were obtained by requiring
that the 662~keV $\gamma$-line from $^{137}$Cs background
must be uniformly distributed along the length of the crystals.
The proportional constants were derived by fixing the
two edges of the distributions to be at 0~cm and 40~cm.
Typical Z-position distributions for events at 662~keV
and 1461~keV evaluated through this prescription 
are shown in Figure~\ref{zpos}. 
The RMS resolution
is 1.3~cm at 3~MeV electron-equivalence, 
based on measurements
with $\alpha$-$\alpha$ cascade events~\cite{uthbkg}.

\begin{figure}
\begin{center}
\includegraphics[width=8cm]{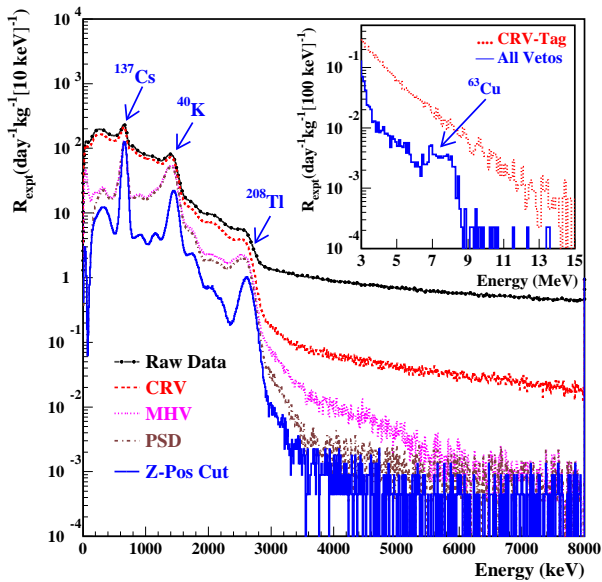}
\caption{
\label{fig_cutspec}
The measured energy spectra 
at various stages of the analysis
showing the effects of successive selection
cuts.
}
\end{center}
\end{figure}

\begin{figure}
\begin{center}
\includegraphics[width=8cm]{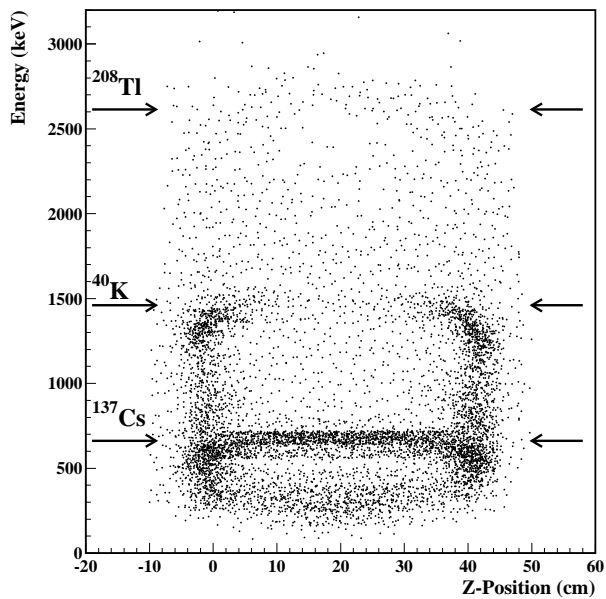}
\caption{
\label{ez2d}
Energy versus Z-position
scatter plot of reconstructed events.
The prominent $\gamma$-lines matching 
the various bands are identified.
}
\end{center}
\end{figure}

The energy is described by:
\begin{equation}
{\rm E} ~ = ~
a_i ~ + ~ b_i \cdot e^{-\alpha_i {\rm Z}} \cdot
\sqrt{Q_{L} \times Q_{R}}  ~~ .
\label{eq_calib}
\end{equation}
The parameters $\alpha_i$'s take into account
possible differences in the attenuation of light transmission
along both directions,
and were fixed by requiring the derived values of E for
the $\gamma$-lines were constant and independent of Z.
The calibration constants $(a_i,b_i)$
were evaluated by a linear fit 
to the $\gamma$-lines. 
The reconstructed energy spectra 
are depicted in Figure~\ref{fig_cutspec}, indicating
RMS resolutions of 5.8\%, 5.2\% and 4.0\% at
$^{137}$Cs, $^{40}$K and $^{208}$Tl $\gamma$-peaks, 
respectively.
A scatter plot of the reconstructed (E,Z) values
for a typical crystal is
shown in Figure~\ref{ez2d}. 
The reconstructed energy of the various bands 
matched well to the corresponding $\gamma$-lines 
within the fiducial volume ($\rm{4 ~ cm < Z < 36 ~cm}$).

\subsection{Event Selection}
\label{sect::evselect}

Neutrino-induced interactions like $\nuebar -$e are of extremely
small cross-section and therefore manifest themselves
as ``single-hit'' (H1) events in only one crystal module
uncorrelated to the rest of the system.
The H1 events were selected from raw data through
selection criteria with
CRV, ``anti-Compton'' multi-hit veto (MHV), and PSD.
The CRV and MHV suppressed cosmic-induced background
and multiple Compton scattering events from ambient
$\gamma$-rays, respectively.
The selected sample is denoted by H1(CRV)
in subsequent discussions.
The $\alpha$ and delay-cascade events from internal
radioactivity~\cite{uthbkg}, as well as 
convoluted events in accidental coincidence,
were identified by PSD.
To minimize background due to ambient $\gamma$-rays,
an internal fiducial volume was defined.
Events at the 
outermost layers of the crystal array were rejected,
and a Z-position cut of 4~cm from both ends 
was applied to the target (inner) crystals.

The various parameters  
in the calibration and selection procedures
were measured before the detector were 
assembled on site.
Typically, about 10\% of the {\it in situ}
data samples uniformly distributed within a
DAQ period were used to provide the
fine adjustments.
Once obtained, the optimal parameters
were applied universally to the rest of the data
set. The energy spectra at the successive stages of candidate
event selection are depicted in 
Figure~\ref{fig_cutspec}.

\begin{table}
\caption{
Summary of the suppression and signal
efficiency factors of successive selection
cuts within the 3$-$8~MeV energy range.
}
\label{tab_cuts}
\begin{ruledtabular}
\begin{tabular}{lcc}
Event Selection & Background & Signal \\
& Suppression & Efficiency \\ \hline
Raw Data & 1.0 & 1.0 \\
Cosmic Ray Veto (CRV)& 0.06 & 0.93 \\
Multi-Hit Veto (MHV) & 0.16 & 0.99 \\
Pulse Shape  & & \\
~ Discrimination (PSD) & 0.34 & $>0.99$ \\
Z-position Cut & 0.36 & 0.80 \\ \hline Combined & 0.0011 & 0.77
\end{tabular}
\end{ruledtabular}
\end{table}

A summary of the background suppression and
signal efficiency factors of the cuts in 
the energy range of interest (3$-$8~MeV)
are summarized in Table \ref{tab_cuts}.
The signal efficiencies 
were derived from the survival fractions
of RT events for the CRV and MHV cuts,
and with the multi-hit Compton events
for the PSD cut. 
The Z-position efficiency
corresponds to a 4~cm cut at both ends
and were accounted for in the
definition of the fiducial volume.

\section{
Background
}

The candidate event selection procedures
of Section~\ref{sect::evselect} resulted in 
a signal-to-background ratio of
about 1/30 at 3~MeV.
The information 
on multiplicity, energy, position, 
cascade event timing and
$\alpha$/$\gamma$ identification
available for every event allowed
the residual background to
be understood, analyzed and suppressed.
In addition,
the Reactor ON/OFF comparisons provided
an independent handle to the background.
These measurements were combined to
improve the background evaluation
which in turn enhanced the experimental
sensitivities.

\subsection{Background Understanding and Diagnostics}
\label{sect::bkgdiag}

Several diagnostic tools which contribute to 
the quantitative understanding of the background 
are discussed in this section.
For completeness, all prominent background channels
are presented, though many of those are below
the physics analysis threshold of 3~MeV. 

\subsubsection{Intrinsic Background}

Measurements of intrinsic radiopurity in the CsI(Tl) crystal
with {\it in situ} data 
were discussed in details in Ref.~\cite{uthbkg}.

The isotope $^{137}$Cs is produced artificially as
fission waste from power reactors and atomic weapon tests. 
Cesium salts are soluble and can easily
contaminate the
raw materials which produce CsI.
The $^{137}$Cs contaminations was measured to be 
${\rm (1.7\pm 0.3) \times 10^{-17}~ g/g}$,
and were uniform across the length of the crystals, 
as depicted in Figure~\ref{zpos}. 

The cascade events provided measurements on the 
naturally-occurring 
$^{238}$U, $^{232}$Th and $^{235}$U series,
which were
${\rm (0.82\pm 0.02) \times 10^{-12} ~ g/g}$, 
${\rm (2.23\pm 0.06) \times 10^{-12} ~ g/g}$ and 
${\rm <4.9 \times 10^{-14} ~ g/g}$, respectively,
assuming secular equilibrium. 
The $\beta$-decays of $^{208}$Tl followed by $\gamma$'s in 
coincidence could in principle lead to background in
the signal region.
From the measured level of $^{228}$Th in the
target, the contribution of this background
at 3$-$5~MeV was evaluated to be only $\alt 11\%$ of the 
expected $\nuebar -$e signals. 

In addition, trace admixtures of 
the fission daughter $^{129}$I and of
the naturally-occurring $^{40}$K 
in the raw CsI powder were measured with
accelerator mass spectrometry techniques
to be $< 1.3 \times 10^{-13} ~ {\rm g/g}$ and
$< 2 \times 10^{-10} ~ {\rm g/g}$,
respectively~\cite{texonoams}. 
Neutron capture on $^{133}$Cs produced $^{134}$Cs 
at the level of $\sim 5 \times 10^{-20} ~ {\rm g/g}$,
as measured with the {\it in situ} two-hit
background discussed in Section~\ref{sect::h2bkg}. 

\subsubsection{Ambient Radioactivity}

The H1 spectra of Figure~\ref{fig_cutspec}
show several $\gamma$-lines, the most prominent
ones were those from $^{137}$Cs,
$^{40}$K and $^{208}$Tl.
The background dropped by several orders of magnitude
beyond the natural radioactivity end-point of 2.6~MeV.
The cut-off at 8~MeV 
corresponded to the end-point of 
$\gamma$-rays emissions
following neutron capture.
The lines are crucial for energy calibration, 
system stability monitoring, and
background diagnostics.
Apart from $^{137}$Cs which
is an intrinsic radioactivity, 
the other sources are external to the CsI(Tl) target.
Distributions of the Z-position
were heavily attenuated from
the edge of the crystals, as illustrated for
the case of $^{40}$K in Figure~\ref{zpos}.

\subsubsection{Cosmic Ray Tagging Efficiency}

The cosmic-ray
tagging efficiency ($\epsilon_{\mu}$)
is the probability that the cosmic-ray induced events
actually produce a ``cosmic-ray tag'' (CRT).
The inefficiency ($1 -\epsilon_{\mu}$)
was due to incomplete geometrical coverage
and the 
light collection deficiencies of the
large-area scintillator panels.

High energy events above the end-point
of (n,$\gamma$) energy scale 
of about 8~MeV 
are all cosmic-ray induced.
These events provided a clean sample
for $\epsilon_{\mu}$
to be measured. For improved experimental
control, three-hit events (H3) 
between 8$-$14~MeV were selected, and
$\epsilon_{\mu}$ is given by
\begin{equation}
\epsilon_{\mu}  ~  = ~
{\rm \left[ ~ \frac{H3 ( CRT )}{H3 (Total) } ~ \right]  }
\label{eq::creff}.
\end{equation}
The  $\epsilon_{\mu}$ averaged over
all the DAQ periods was measured to be 92\%,
while the variations among periods were less than
1\%.

\subsubsection{Diagnostics of Two-Hit Background}
\label{sect::h2bkg}

Multi-hit events were unrelated to neutrino interactions
and therefore provided unambiguous diagnostics to the
background sources. 
Displayed in Figure~\ref{2hspec} is a
scatter plot of two-hit (H2) events
after CRV cut.
Several features were noted which revealed
the nature and locations of the dominant background
sources, discussed as follows: 

\begin{figure}
\begin{center}
\includegraphics[width=8cm]{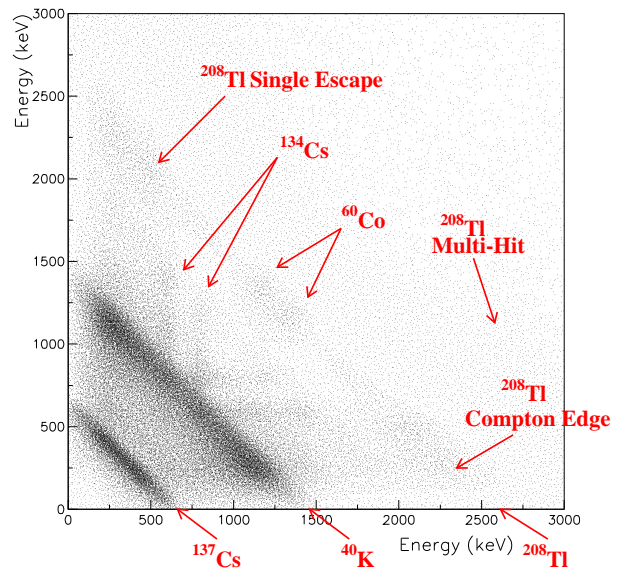}
\caption{\label{2hspec} 
Scatter plot of H2 events 
after cosmic-ray veto in 0$-$3~MeV,
showing bands on 
$^{60}$Co, $^{134}$Cs and $^{208}$Tl single-escape,
as well as the
correlated $\gamma$'s from $^{208}$Tl.
}
\end{center}
\end{figure}

\begin{description}

\item [(i) $^{208}$Tl Induced Pair Production:]
The single escape peak following pair production
of the $^{208}$Tl 2614~keV $\gamma$'s can be identified.
The Z-position distribution of these events 
confirmed that the sources were external to the target. 
As discussed in Section~\ref{sect::bkgevaluation},
pair production events are crucial for  
background evaluation because of their
distinctive topologies.

\item [(ii) $^{60}$Co Contaminations:]
It was established that the dominant 
reactor-induced radioactivity in KSNL was
$^{60}$Co which existed as dust 
in the laboratory area~\cite{texonomunu}.
Their contributions varied 
between DAQ periods
due to different levels of contaminations during
hardware installation. 
Events due to the correlated $\gamma$'s at
energy 1173~keV and 1332~keV 
from $^{60}$Co 
can be located in Figure~\ref{2hspec}.
They were uniformly distributed along the Z-position,
signifying that some $^{60}$Co dust got into the
target volume between crystals during installation.
The  measured contamination level is 
$3 \times 10^{-20} ~ {\rm kg^{-1}}$.
However, the total energy of the $^{60}$Co lines
is below the $3 - 5 ~ {\rm MeV}$ signal region
relevant to this measurement.

\item [(iii) Neutron Capture Induced $^{134}$Cs:]
Trace amount of $^{134}$Cs 
(${\rm \tau_{1/2}=2.05 ~ yr}$ ; ${\rm Q = 2.06 ~ MeV}$)
was produced by
neutron capture on $^{133}$Cs
within the CsI(Tl) target.
It decays via $\beta$-emission together with
two $\gamma$'s of energy 605~keV and
796~keV in coincidence.
These events were tagged in the H2
plot in Figure~\ref{2hspec}.
The intensity 
distribution is uniform over the length
of the crystals, verifying the sources were internal.
The measured contamination level is 
$5 \times 10^{-20} ~ {\rm g/g}$.
The Q-value is below the physics analysis threshold and
hence these decays would not contribute to the
background of this measurement.

\item [(iv) Cascade $\gamma$-rays from $^{208}$Tl:]
Decays of $^{208}$Tl 
are characterized by several $\gamma$-rays
emitted in cascade. 
Coincidences of $\gamma$-rays
at 510, 583 and 860~keV 
with the prominent line of 2614~keV
can be identified in the H2 scatter 
plot of Figure~\ref{2hspec}.
The evaluation of the contributions
of this channel to H1 events 
is crucial to background suppression,
and is addressed
in Section~\ref{sect::tl208bkg}.

\item [(v) Neutron Capture on $^{63}$Cu:]
The main shielding materials in the vicinity
of the target were OFHC copper.
Neutron (n,$\gamma$) capture on $^{63}$Cu
has relatively large cross-section (4.5 b),
giving rise to
high energy $\gamma$'s at
7637~keV and 7916~keV.
These were observed in
H1 spectrum shown in the inset of
Figure~\ref{fig_cutspec}.

\end{description}

\subsubsection{Pair Production Event Samples}

Pair production background manifested themselves mostly
as three-hit events (${\rm H3_{PP}}$).
They were selected by requiring two crystals each 
having 511~keV of energy back-to-back to the third one.
These samples have distinctive topology not
contaminated by other background channels.
Coupled with the known energy dependence
of the pair production cross-section,
these samples provided measurements of
the {\it in situ} $\gamma$-spectra,
and therefore 
were crucial for subsequent background evaluation.

\begin{figure}
\begin{center}
\includegraphics[width=8cm]{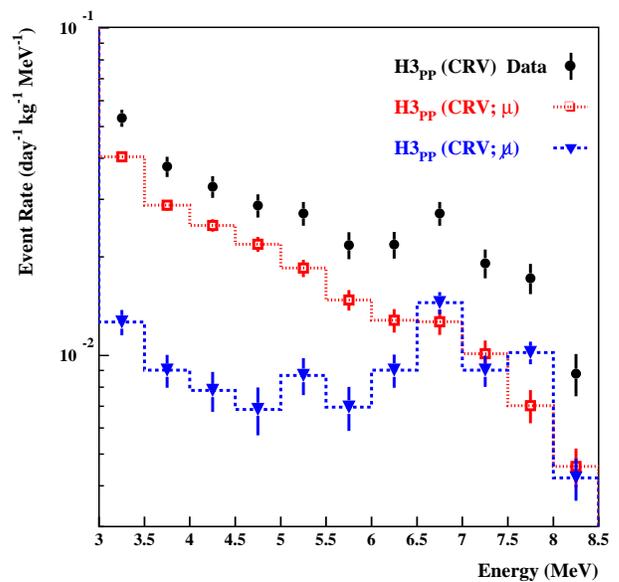}
\caption{\label{h3pp}
Three-hit pair production spectra for
CRV events, and further differentiating
into the cosmic-ray related and 
unrelated components. 
After scaling with the known pair production
cross-sections, these spectra provide
the {\it in situ} $\gamma$-ray background 
at the detector.
}
\end{center}
\end{figure}

The ${\rm H3_{PP}}$ spectrum for CRV events 
are displayed in Figure~\ref{h3pp}. 
This was produced by 
the $\gamma$-ray background whose 
contributions to the H1(CRV) signals
were evaluated.
There were two components to
this high energy $\gamma$-ray background:
\begin{description}
\item [(i)] cosmic-ray induced events with missing CR-tags,
whose rates are given by
\begin{equation}
\label{eq::h3pp}
\rm{
H3_{PP} ( CRV ; \mu ) ~ = ~ 
\left[  \frac{ ( 1 - \epsilon_{\mu} ) }{ \epsilon_{\mu} } \right]
H3_{PP} ( CRT )  ~ ,
}
\end{equation}
where $\epsilon_{\mu}$ is the cosmic-ray tagging efficiency
measured with Eq.~\ref{eq::creff},
and
\item [(ii)] ambient radioactivity unrelated to cosmic-rays, which 
can be evaluated with
\begin{equation}
\label{eq::h3ppnomu}
\rm{ 
H3_{PP} ( CRV ; \nearrow \hspace*{-0.4cm} \mu ) 
 ~ =  ~ 
H3_{PP} (CRV) - H3_{PP} (CRV;\mu)  ~~ ,
}
\end{equation}
also depicted in Figure~\ref{h3pp}.
\end{description}

\subsection{Background Evaluation}
\label{sect::bkgevaluation}

The experiment 
focused on the ${\rm 3 ~ MeV < T < 8 ~ MeV}$ energy
range as the physics analysis window.
The $\nuebar -$e signal region 
is expected to be at 3$-$5~MeV
due to rapid decrease of the reactor $\nuebar$-spectra.

The background diagnostics in Section~\ref{sect::bkgdiag}
demonstrated that convoluted $\gamma$-rays 
from $^{208}$Tl, cosmic-ray events
with missing CRV tags, 
as well as ambient high energy photons
could contribute to the H1 background [H1(BKG)].
The experimental design allowed
quantitative measurement of these 
background which resulted in the 
extraction of the $\nuebar -$e signal
events with good accuracy.

\begin{figure}
\begin{center}
\includegraphics[width=8cm]{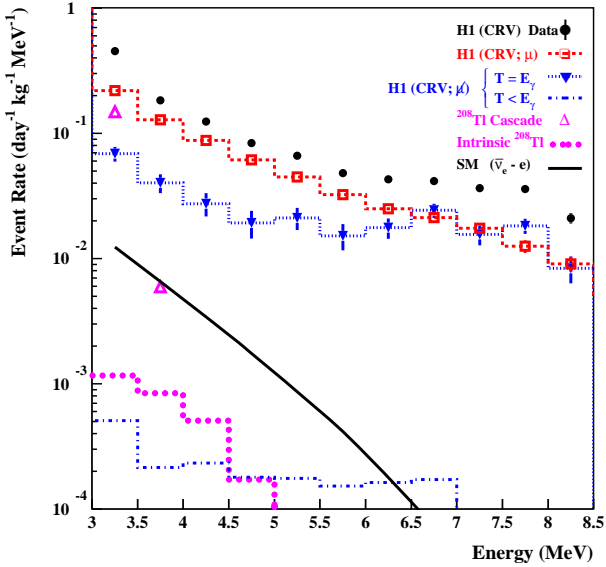}
\caption{\label{bkgchannel}
Measured H1 spectrum and
the different background channels
evaluated in Section~\ref{sect::bkgevaluation}. 
The SM $\nuebar -$e contributions are
overlaid.
}
\end{center}
\end{figure}

The evaluation of the various
background channels is discussed in the
following sub-sections. 
Their contributions 
are depicted in Figure~\ref{bkgchannel},
where 
the expected SM $\nuebar -$e spectrum
is overlaid for comparison.

\subsubsection{Cascade $\gamma$-Rays from $^{208}$Tl}
\label{sect::tl208bkg}

Decays of $^{208}$Tl are followed by emissions
of $\gamma$-rays in coincidence, 
having energy
${\rm E_{Tl} (1,2,3,4)}$= 
2614.5, 860.56, 583.2 and
510.8~keV, 
and at intensity ratios of
99, 12.8, 86.2 and 25\%
per $^{208}$Tl-decay,
respectively.
Two-fold coincidence manifested as 
H2 events were identified 
in the scatter plot of
Figure~\ref{2hspec}.
Events with both $\gamma$-rays 
hitting and depositing all
energy in the same crystal
would become H1 background to
the $\nuebar -$e signals.

\begin{figure}
\begin{center}
\includegraphics[width=8cm]{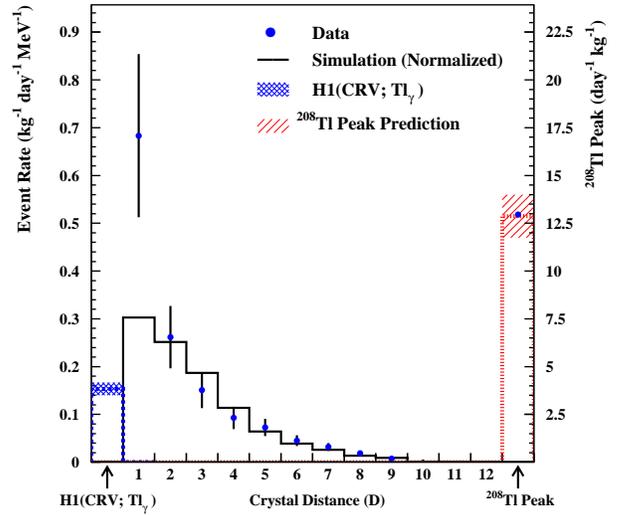}
\caption{
\label{fig_tlpred}
Comparison of measured H2 correlated events 
with simulation results on the cascade
$\gamma$-rays due to $^{208}Tl$ decays.
The entry 
$\rm{H1 ( CRV ; Tl_{\gamma} ) }$
denotes the predicted 
H1 background produced by two $\gamma$'s 
depositing all energy only in the same crystal.
The measured strength of the $^{208}$Tl
peak provided consistency cross-checks to  
the simulations.
}
\end{center}
\end{figure}

The probabilities were studied by full 
simulations with GEANT software packages~\cite{geant3},
incorporating realistic angular correlations and 
branching ratios for the
$^{208}$Tl decays~\cite{tl208angle}.
The sources were located
at the PMTs and their voltage dividers, 
which were the only materials 
other than OFHC copper and teflon in the vicinity
of the target.

The measured H2 distribution
of ${\rm E_{Tl}(1) \oplus E_{Tl}(2,3,4)}$ 
as a function of distance between the two crystals (D)
was displayed in Figure~\ref{fig_tlpred}.
The simulation results are overlaid, 
the normalization of which was fixed by best-fit 
to the H2 distribution for events with
separation more than one crystal 
(D$>$1).
Excellent agreement with the strength of 
the  ${\rm E_{Tl}(1)}$ single-$\gamma$ 
H1 peak at 2614~keV
was demonstrated. 
It served as important consistency check
and tools for systematic studies.
The data point at D=1 
denotes H2 events with hits from adjacent
crystals.
The measured intensity was significantly
larger than the expected
contributions from 
correlated $\gamma$'s due to $^{208}$Tl. 
The excess was attributed to
multiple Compton scatterings 
at adjacent crystals
from a single high energy photon.
This was reproduced in simulations
studying H2 events with single photons.

The entry at D=0,
denoted by $\rm{H1 ( CRV ; Tl_{\gamma} ) }$,
corresponds to
the prediction of the H1 events
having the two $\gamma$'s depositing energy 
exclusively in the same crystal.
It was adopted for subsequent background subtraction. 
The relative intensities to the 2614~keV reference peak,
expressed as ratios of
${\rm \left[ E_{Tl}(1) \oplus E_{Tl}(N) \right] / E_{Tl}(1) } $ 
in H1-events, 
are 0.13\%, 0.33\%, and 0.16\%,
for N=2,3,4, respectively.

\subsubsection{Cosmic-Ray Induced Background} 

Once the cosmic-ray tagging efficiency ($ \epsilon_{\mu}$)
was measured with Eq.~\ref{eq::creff},
the comic-ray induced H1 background with missing CR tags 
was derived using
\begin{equation}
\label{eq::crpred}
\rm{
H1 ( CRV ; \mu ) ~ = ~ 
\left[  \frac{ ( 1 - \epsilon_{\mu} ) }{ \epsilon_{\mu} } \right]
H1 ( CRT )  ~ ,
}
\end{equation}
similar to the $\rm{H3_{PP}}$ channel in Eq.~\ref{eq::h3pp}.

\subsubsection{Ambient $\gamma$-Ray Background }

This background channel 
$\rm{ H1 ( CRV ; \nearrow \hspace*{-0.4cm} \mu )}$ 
is due to ambient high energy photons
emitted mostly through thermal neutron capture
by the surrounding materials.
This was derived directly through the
$\rm{ H3_{PP} ( CRV ; \nearrow \hspace*{-0.4cm} \mu )}$
spectrum of Figure~\ref{h3pp}. 
The sharp cut-off at $\sim$8~MeV indicates the dominance
of (n,$\gamma$) processes.

The background 
can be further divided into two
categories, according to the different methods of evaluation.

\begin{description}

\item[(i) Full Energy Deposition ($T = E_{\gamma}$):]
The $\gamma$-rays lose all its energy within a single
crystal through multiple Compton scatterings or 
pair production with both annihilation photons fully absorbed.
The rate
is given by
\begin{equation}
\rm{
  H1 ( CRV ; \nearrow \hspace*{-0.4cm} \mu )
~ = ~ 
\left[
\frac{ H3_{PP} ( CRV ; \nearrow \hspace*{-0.4cm} \mu ) }
{H3_{PP} (CRT)} \right]
\cdot H1 ( CRT ) ~~. 
}
\label{eq::h1crvnomu}
\end{equation}
The evaluation of 
$\rm{H3_{PP} ( CRV ; \nearrow \hspace*{-0.4cm} \mu ) }$
followed from Eq.~\ref{eq::h3ppnomu}.

\item[(ii) Partial Energy Deposition ($T < E_{\gamma}$):]
The ambient $\gamma$'s 
could undergo single Compton scattering 
after which the outgoing photons left the detector
without further interactions.
Only a fraction of the incident energy
would be deposited in a single crystal.
This background channel was studied with
full-scale simulations
using the {\it in situ} cosmic-unrelated
$\rm{H3_{PP} ( CRV ; \nearrow \hspace*{-0.4cm} \mu ) }$
spectrum of Figure~\ref{h3pp}
for flux normalization.
For consistency check,
the strength of the H1 full energy ($T = E_{\gamma}$)
spectra of Figure~\ref{bkgchannel}
was successfully reproduced.
The contribution by this channel 
to H1(BKG) at 3$-$5~MeV was 
only $\alt$5\%  
of the expected SM 
$\nuebar -$e signals.
\end{description}

\subsubsection{Combined Evaluation}
\label{sect::combinedbkg}

It can be derived from Figure~\ref{bkgchannel} 
that $\agt$99\% of
the H1(CRV) events
can be accounted for by 
the $\nuebar -$e signals 
as well as the three
dominant background channels,
such that: 
\begin{eqnarray}
& & 
{\rm H1(CRV) = H1 ( \nuebar - e ) + H1 (BKG) } ~~ ;  \nonumber \\ 
& & ~~ {\rm H1(BKG) }
  \cong  \\
& & ~~~~~~ {\rm H1 ( CRV ; Tl_{\gamma} ) } {\rm +
H1 ( CRV ; \mu )  +
  H1 ( CRV ; \nearrow \hspace*{-0.4cm} \mu ) ~~ , \nonumber
}
\label{eq::sumh1crv}
\end{eqnarray}
where the three contributions are given by
Fig.~\ref{fig_tlpred} and 
Eqs.~\ref{eq::crpred}\&\ref{eq::h1crvnomu}, respectively.
The sub-dominant terms include
intrinsic radiopurity and ambient $\gamma$-ray background
with partial energy deposition which contributed
at the $\alt$0.5\% level of H1(BKG).

The $\rm{ H1 ( CRV ; Tl_{\gamma} )}$ channel
was important only in the 3$-$3.5~MeV energy bin.
The other two channels due to 
high energy $\gamma$ interactions
were dominant over the
entire energy range of interest.
Their combined contributions 
were simplified by 
Eqs.~\ref{eq::h3pp},\ref{eq::h3ppnomu},\ref{eq::crpred}\&\ref{eq::h1crvnomu},
to become:
\begin{equation}
{\rm H1 ( CRV ; \mu ) } + 
{\rm H1 ( CRV ; \nearrow \hspace*{-0.4cm} \mu ) }
= 
{\rm \left[
\frac{ H3_{PP} ( CRV ) }
{H3_{PP} (CRT)} \right] } 
\cdot {\rm H1 ( CRT ) } ~~ . 
\label{eq::net}
\end{equation}
That is, the dominant contribution to
${\rm H1(BKG)}$ was related to the $\rm{H3_{PP}}$ sample
through a simple ratio of events with and without
CR-tags.

\section{Systematic Uncertainties}

A summary of the sources of systematic errors 
[$\delta_{sys} {\rm ( Source)}$] and
their contributions to the measured  $\xi$-ratio
[${\rm \Delta_{sys}  ( \xi ) }$] 
is given in Table~\ref{syserr}.
An uncertainty of 3\% was adopted for 
the evaluation of 
the high energy reactor $\nuebar$-spectra.
The signal efficiencies for the selection procedures
discussed in Section~\ref{sect::evselect} 
were accurately measured
with high statistics using the RT events.
The fiducial mass uncertainties originated
from the Z-position resolution of 1.3~cm.

\begin{table}
\caption{\label{syserr}
Summary of the sources of systematic errors 
[$\delta_{sys} {\rm (Source)}$] and their contributions to
the measurement uncertainties [$\Delta_{sys} ( {\rm \xi} )$].
The various components to the signal strength
are summed,
while those to the background subtraction
are averaged.
}
\begin{ruledtabular}
\begin{tabular}{lcr}
Sources & $\delta_{sys}$(Source) & $\Delta_{sys} ( {\xi } )$ \\ \hline
\multicolumn{3}{l}{\underline{Signal Strength} :} \\
~~$\Phi _{\nu}$ Evaluation & $<$3\%  & $<$0.03 \\
~~Efficiencies for Neutrino Events & $<$1.3\% & $<$0.013 \\
~~Fiducial Target Mass & $<$4\% & $<$0.04 \\
$\ast$ Combined (Signal) & $-$  & $<$0.052 \\ \hline
\multicolumn{3}{l}{\underline{Background Subtraction} :} \\
~~Reactor OFF Measurement & $< 0.4\%$ & $<$0.06 \\
~~Background Evaluation &  & \\
~~~~$\rm{\odot H1(CRV;Tl_{\gamma})}$  & $<$3\% & $<$0.08 \\ 
~~~~$\rm{\odot H1(CRV;\mu) + H1 ( CRV ; \nearrow \hspace*{-0.4cm} \mu ) }$
& $<1\%$ & $<$0.17 \\ 
~~~~~ Net & $-$ & $<$0.19 \\
$\ast$ Combined (Background) & $-$  & $<$0.15 \\ \hline
Total & & $<$0.16
\end{tabular}
\end{ruledtabular}
\end{table}

The systematic effects on 
background evaluation were studied with
event samples unrelated
to neutrino interactions accumulated over all DAQ periods.
These include data from the Reactor OFF periods as well as 
those with energy above the 8~MeV end-point of the 
reactor neutrino spectra.
Individual methods were demonstrated to be able
to account for the neutrino-unrelated
background to certain accuracy levels, which
were in turn assigned as the systematic uncertainties 
of those methods.

\begin{description}
\item [(i)] {\bf Reactor ON/OFF Comparison:}
The intensity of the $^{208}$Tl $\gamma$-line allowed
the stability of the hardware systems to be monitored
and demonstrated to good statistical accuracies. 
The window within 3$-$8~MeV at
the Reactor OFF periods consisted exclusively of
background  and provided an additional monitor.
The stability of the measured intensities
of the $^{208}$Tl $\gamma$-line at 2614~keV 
in Period III relative to the whole-period average 
is illustrated in Figure~\ref{fig_system}a. 
Summary of all results are tabulated in Table~\ref{stability}.
The good reduced-$\chi ^2$ ($\chi^2$/dof)
indicate the data were stable within individual periods.
The hardware instability level demonstrated with the combined 
data is $< 0.4\%$. 

\begin{table}
\caption{\label{stability}
Stability levels ($\delta_{\pm}$)
of the various neutrino-unrelated
channels.
}
\begin{ruledtabular}
\begin{tabular}{ccc}
Channels/Period & $\chi ^2$/dof & $\delta_{\pm}$ (\%)  \\ \hline
\multicolumn{3}{l}{\underline{$^{208}$Tl Intensity} :} \\
II  & 19/13 & 0.85 \\
III  & 38/43 & 0.61 \\
IV  & 33/27 & 0.81 \\
V & 8.4/8 & 0.91 \\
\multicolumn{3}{l}{\underline{Reactor OFF 3$-$8 MeV H1(CRV) Rates} :} \\
II  & 15/14 & 3.18 \\
III  & 11/11 & 3.51 \\
IV  & 8.1/8 & 3.60 \\
V & 7.2/5 & 3.22 \\ \hline Combined & $-$ & $<0.4 \%$
\end{tabular}
\end{ruledtabular}
\end{table}

\begin{figure}
\begin{center}
{\bf (a)}\\
\includegraphics[width=8cm]{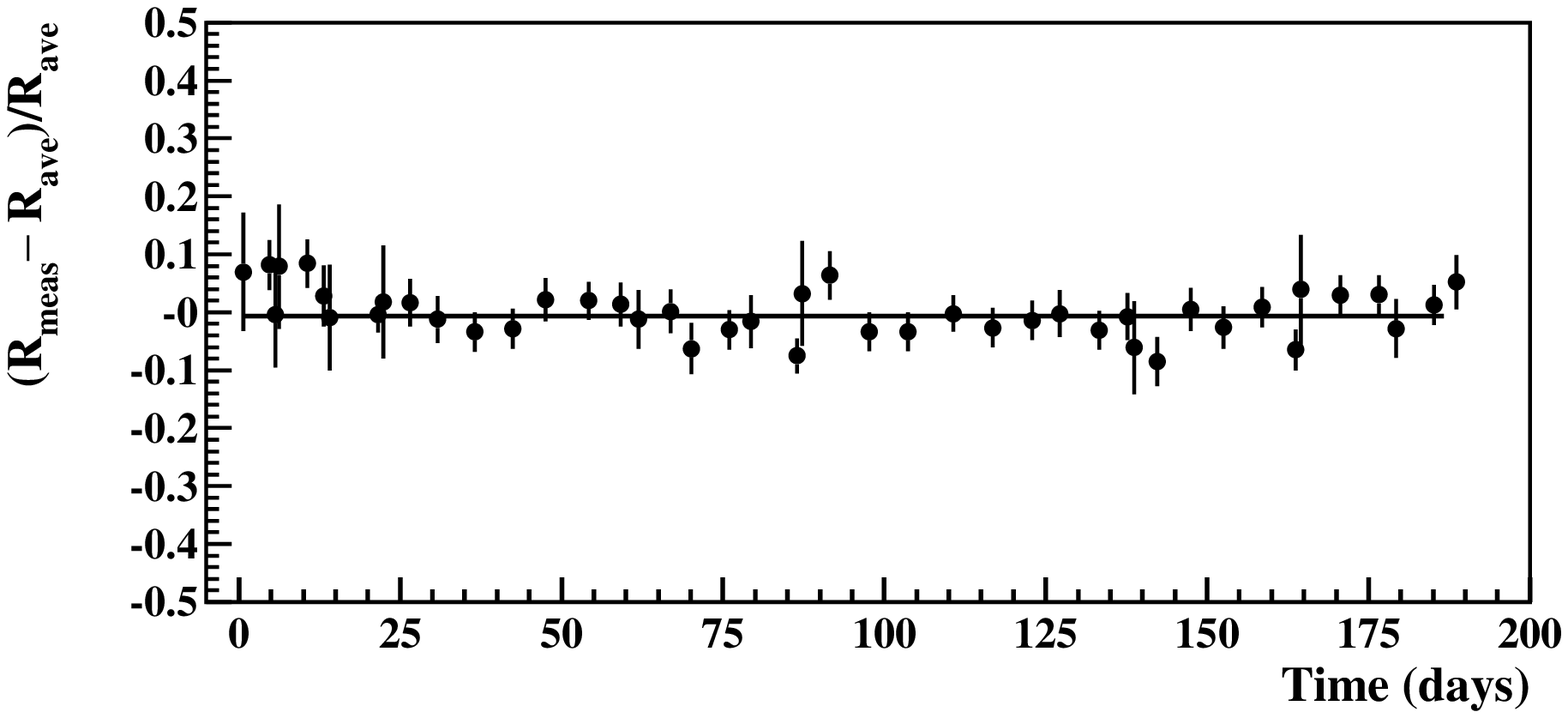}\\
{\bf (b)}\\
\includegraphics[width=8cm]{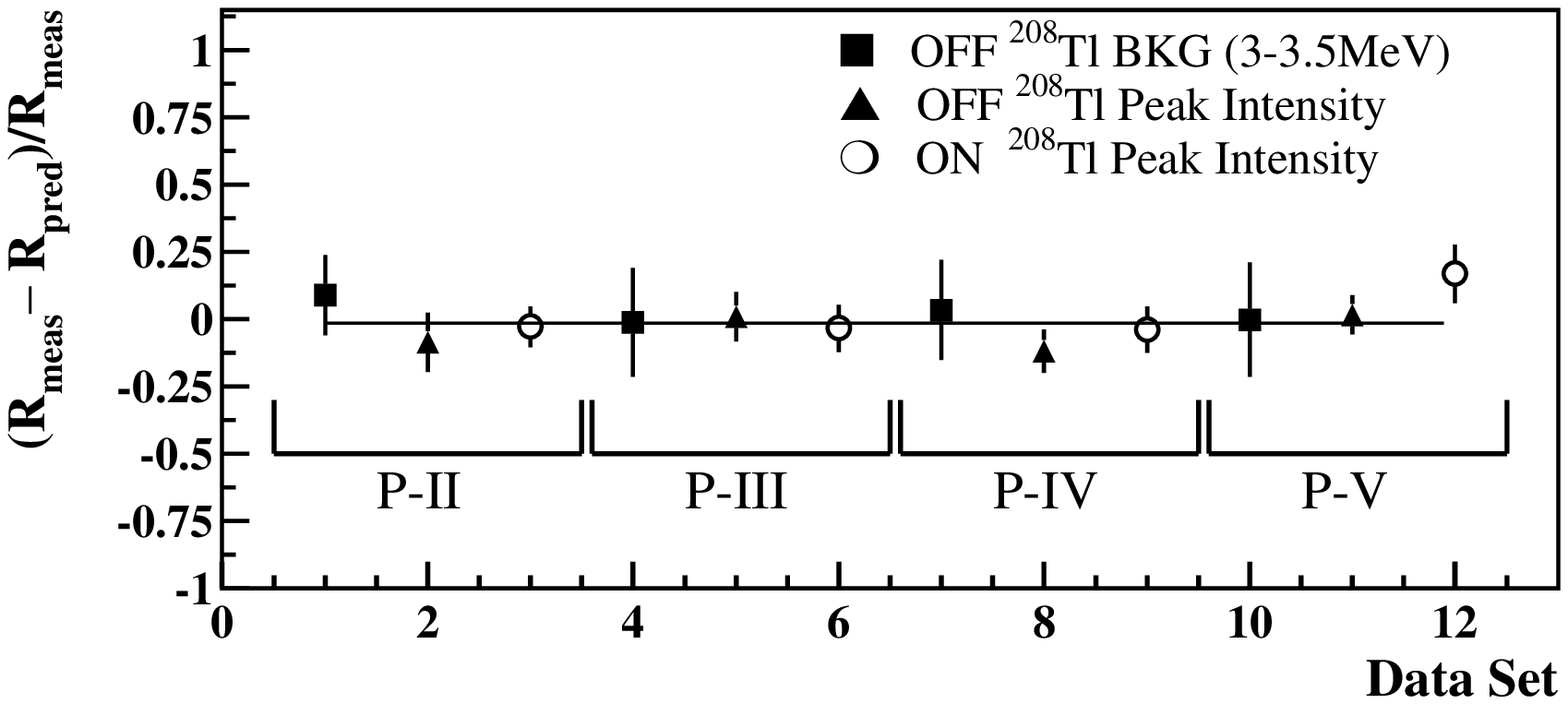}\\
{\bf (c)}\\
\includegraphics[width=8cm]{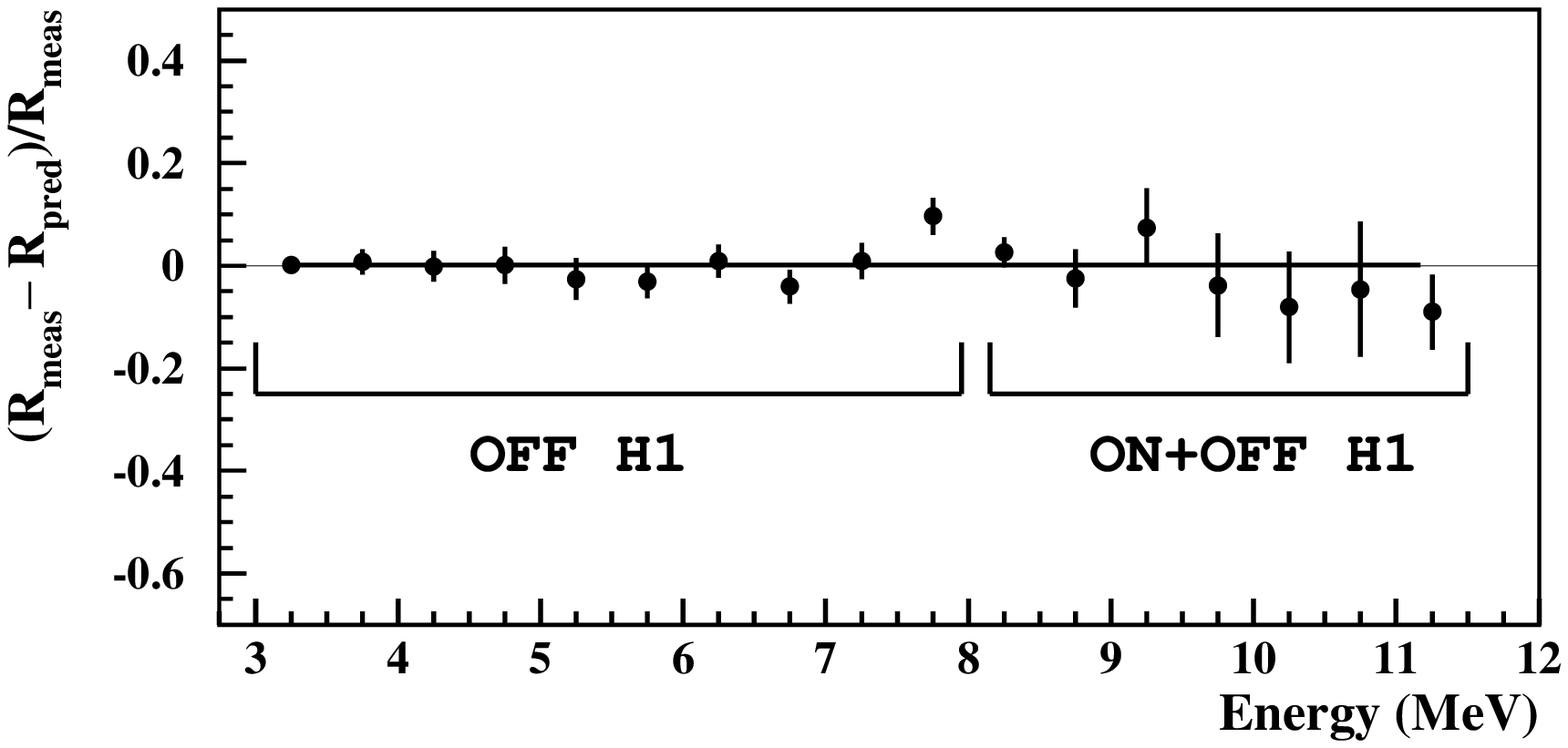}\\
\caption{
\label{fig_system}
Data from which the systematic uncertainties were derived:
(a) stability of the $^{208}$Tl peak intensities in Period III;
(b) comparisons of $^{208}$Tl intensities  
evaluated by simulations with measured background on
neutrino-unrelated data samples,
and
(c) comparisons of the evaluated 
$\rm{ H1 (CRV;\mu) + H1 ( CRV ; \nearrow \hspace*{-0.4cm} \mu ) }$
background channels with measured background on
neutrino-unrelated data samples.
}
\end{center}
\end{figure}

\item
[(ii)] {\bf $^{208}$Tl Induced $\gamma$-Ray Cascade Background:}
The simulation software and the normalizations discussed
in Section~\ref{sect::tl208bkg} were cross-checked by applying them
to compare with the measured intensities of the
2614~keV $\gamma$-line following
$^{208}$Tl decays for all periods,
and with the H1 events at 3$-$5~MeV for only Reactor OFF.
The relative deviations
between the measured and predicted rates
${\rm [  ( R_{meas} - R_{pred} ) / R_{meas} ] }$
are depicted in Figure~\ref{fig_system}b,
showing consistency with zero 
$[ = (-0.013 \pm 0.029 )  ~ {\rm at }  ~ \chi ^2/{\rm dof} = 6/11 ]$.
The fitting error of 3\% represents an upper bound
of the systematic uncertainties to the
$\rm{ H1 ( CRV ; Tl_{\gamma} ) }$ 
background component.

\item
[(iii)] {\bf Dominant 
${\rm \bf H1 ( CRV ; \mu ) + 
 H1 ( CRV ; \nearrow \hspace*{-0.4cm} \mu ) }$
Background:}
The evaluations of the 
${\rm H1 ( CRV ; \mu ) + 
 H1 ( CRV ; \nearrow \hspace*{-0.4cm} \mu ) }$
combined contributions
in Eq.~\ref{eq::net}
were cross-checked 
with measurements on 
neutrino-unrelated samples  
at 3$-$8~MeV from the Reactor OFF periods 
and at 8$-$12~MeV
from both ON/OFF periods.
The relative deviations 
${\rm [ ( R_{meas} - R_{pred} ) / R_{meas} ] }$
were consistent with zero 
$[ = ( 0.0021 \pm 0.0081 ) ~ {\rm at } ~ 
\chi ^2/{\rm dof} = 14.5/16 ]$,
as illustrated in Figure~\ref{fig_system}c for
the combined data set.
The fitting error of 1\% represents an upper bound
of the systematic uncertainties. 

\end{description}

The contributions of the individual systematic 
effects to the $\nuebar -$e cross-section 
measurement were then derived.
The various $\delta_{sys}$(Source) terms
were imposed on the data, and
the changes introduced on $\xi$
were the corresponding systematic
uncertainties $\Delta_{sys} ( \xi )$
listed in Table~\ref{syserr}.

\section{Physics Results}
\label{sect::results}

Intermediate results of the experiment were previously
reported~\cite{ichep08}. 
The final physics results presented in 
this section are based on the complete data set, 
and include
contributions from systematic uncertainties, as well as
improvements in the background evaluation.

\subsection{Formulation}

The experimentally measured rates for neutrino events 
$[ R_{expt} ( \nu ) ]$ in Eq.~\ref{eq::xidef}
are given by:
\begin{equation}
R_{expt} ( \nu ) 
=
R_{\rm H1} ({\rm ON)} 
-
 R_{\rm H1} ({\rm BKG}) ~~ ,
\label{eq_sys}
\end{equation}
where $R_{\rm H1} (\rm ON)$ is the measured 
H1(CRV) spectra for Reactor ON data, and
$R_{\rm H1} ({\rm BKG})$ is the background
derived from the statistical average of
two different measurements: 
(1) Reactor OFF data, and
(2) sum of the dominant and sub-dominant
background contributions to ${\rm H1(BKG)}$
in both the Reactor ON and OFF periods,
the derivations of which are
discussed in Section~\ref{sect::combinedbkg}.

Data from the four independent DAQ periods 
were used, combining to give
a total of 29882(7369)~kg-day of 
fiducial mass exposure during Reactor ON(OFF), respectively.
The adopted analysis window is 3$-$8~MeV
spread out uniformly over $N_{bin}=10$ energy bins.

\subsection{Cross-Section}

The cross-section ratio $\xi$ defined in Eq.~\ref{eq::xidef}
was derived with a
minimum-$\chi ^2$ fit, defined by
\begin{equation}
\chi ^{2}=\sum_{i=1}^{N_{bin}}
\left\{ 
\frac{ [ ~  R_{expt} (i) - \xi \cdot R _{SM} (i) ~  ] ^{2}}
{\Delta_{stat} (i) ^2 }  \right\}
~~ ,
\label{eq::chi2}
\end{equation}
where $R_{SM} (i)$ and $R_{expt} (i) $ are SM-expected and 
measured event rates at $i^{th}$ bin, respectively, 
and $\Delta_{stat} (i) $ is the corresponding
statistical error of the measurement. 

\begin{figure}
\begin{center}
{\bf (a)}\\
\includegraphics[width=8cm]{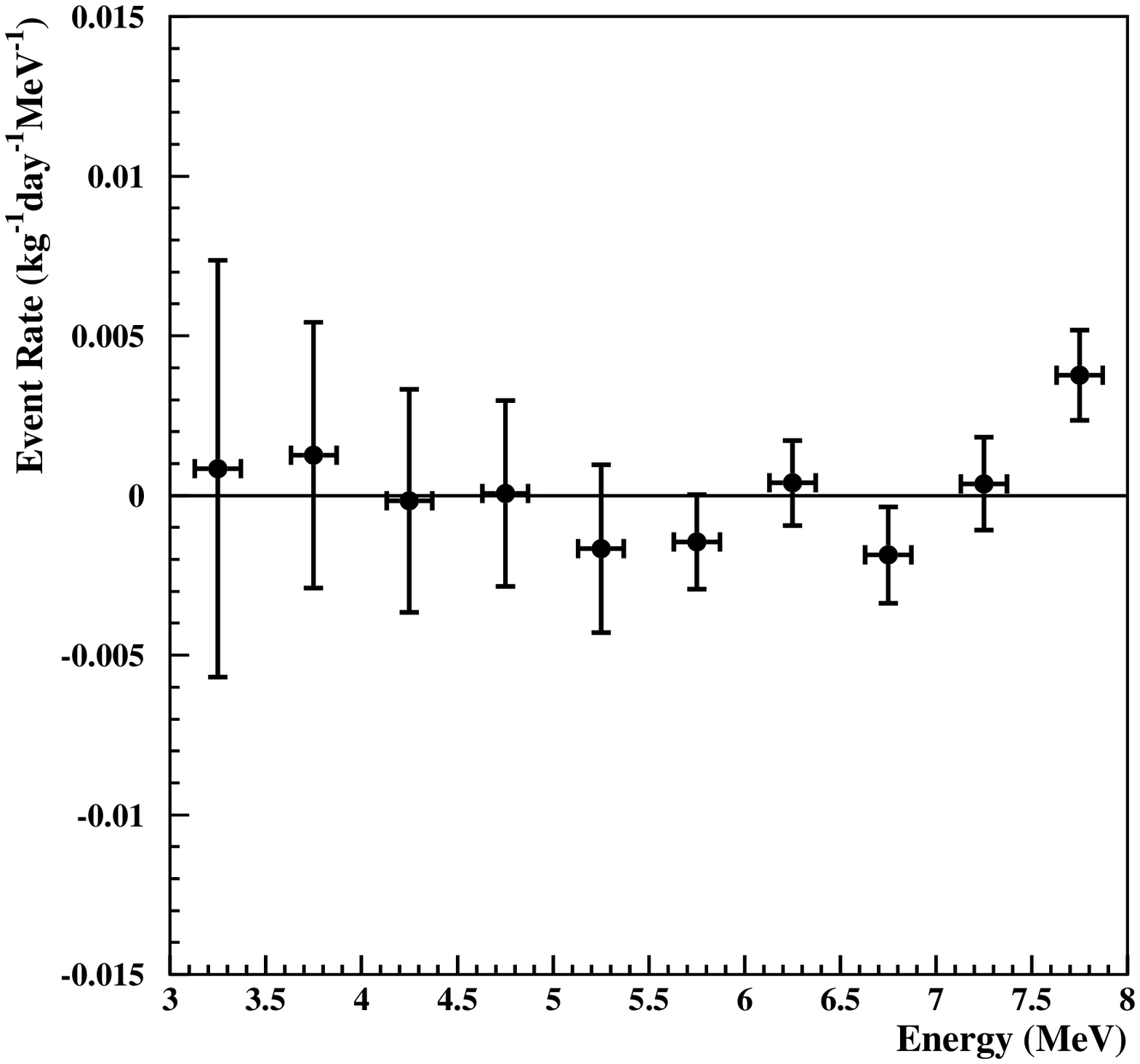}\\
{\bf (b)}\\
\includegraphics[width=8cm]{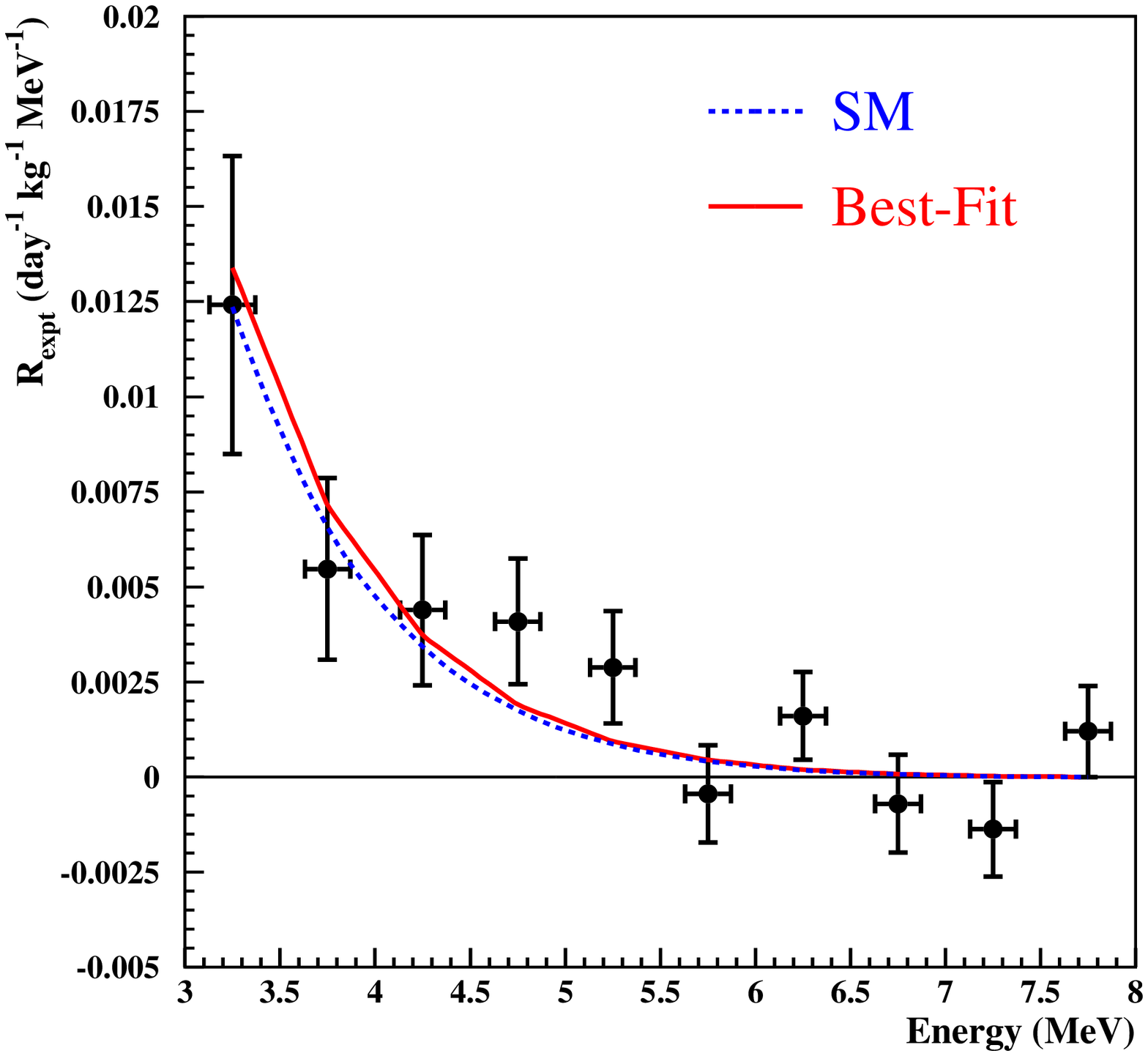}\\
\caption{
\label{residual_fit}
(a)
The residual spectrum  
$\rm{ [ R_{H1} (OFF) - R_{pred} (OFF) ] }$ 
with combined Reactor OFF data.
The best-fit to SM effects is 
consistent with $\xi = 0$. 
(b)
The combined 
residual spectrum 
$[ R_{expt} ( \nu ) = 
R_{\rm H1} ({\rm ON}) - R_{\rm H1} ({\rm BKG}) ]$ 
in the $3-8$~MeV energy region.
The blue and red lines correspond to the SM expectations
and to the best-fit of the data, respectively.
}
\end{center}
\end{figure}

As cross-check, identical procedures were applied
to the combined Reactor OFF data 
$ [ R_{\rm H1} ({\rm OFF}) ] $, 
in which case only
the predicted background 
was subtracted to provide the
residual spectrum displayed in Figure~\ref{residual_fit}a.
Best-fit with Eq.~\ref{eq::chi2} gave 
\begin{equation}
\xi ({\rm OFF}) =  0.03 \pm  0.36 (stat)
\label{eq::xioff}  
\end{equation}
at $\chi ^2$/dof=10.3/9, demonstrating
good overall systematic control of the
background subtraction procedures.

Combining all Reactor ON and OFF data 
from all periods
and adopting the systematic uncertainties 
listed in Table~\ref{syserr},
the ratio
\begin{equation}
\xi = 1.08 \pm 0.21 (stat) \pm 0.16 (sys) 
\label{eq::xionoff}  
\end{equation}
at $\chi ^2$/dof=8.7/9 was derived
following Eq.~\ref{eq::chi2}.
The measured $\nuebar -$e cross-section was
consistent with the SM prediction.

The residual and best-fit spectra
are depicted in Figure~\ref{residual_fit}b.
The $\xi$ ratios derived from individual periods
as well as with background subtraction by
different methods 
are tabulated in Table~\ref{tab_meas}.
As illustrations using
Period II Reactor ON data,
the raw sample consisted of
about $1.94 \times 10^6$ events.
The analysis procedures 
of Section~\ref{sect::evselect}
selected 2074~counts.
A background estimate of
(2016$\pm$17$\pm$8) events was subtracted  
based on the various schemes in Section~\ref{sect::bkgevaluation},
resulting in (57$\pm$27$\pm$8) signal events.
The total $\nuebar$-e sample strength
from all four periods combined
corresponds to 
$[ 414 \pm 80 (stat) \pm 61 (sys) ]$~events.

The consistent distributions of the
best-fit values and their
errors in Table~\ref{tab_meas}
together with the appropriate range of  
the $\chi^2$/dof values
indicate robustness 
of the analysis procedures.
These results represent a probe to SM at 
$Q^2 \sim ( 3 \times 10^{-6} ) ~ {\rm GeV^2}$ 
and improve over those from 
previous reactor neutrino 
experiments~\cite{savannah,kras,rovno,munu}.

\begin{table}
\caption{
Summary of the measured values 
of $\xi$ and $\chi ^2$/dof 
over individual DAQ periods,
as well as
with the different background
subtraction schemes in the total data set.
}
\label{tab_meas}
\begin{ruledtabular}
\begin{tabular}{lcc}
& $\xi$  & $\chi ^{2}$/dof  \\ \hline
\multicolumn{3}{l}{\underline{Individual Period} :}\\
~~ II & $1.15\pm 0.55 \pm 0.17$ &  $8.5/9$ \\
~~ III & $1.03\pm 0.43 \pm 0.20$ &  $8.3/9$ \\
~~ IV & $1.18\pm 0.36 \pm 0.19$ &  $7.3/9$ \\
~~ V & $0.97\pm 0.42\pm 0.20$ &  $9.9/9$ \\ \hline
\multicolumn{3}{l}{\underline{All Periods Background Subtraction} :}\\
~~ Measurement Reactor OFF & $1.25 \pm 0.43 \pm 0.08$ & $7.4/9$ \\
~~ Evaluation Reactor OFF & $1.33 \pm 0.37 \pm 0.22$ & $6.9/9$ \\
~~ Evaluation Reactor ON & $0.78 \pm 0.33 \pm 0.20$ & $10.3/9$
\\ \hline Combined & $1.08 \pm 0.21 \pm 0.16$ & $8.7/9$
\end{tabular}
\end{ruledtabular}
\end{table}

\subsection{Electroweak Parameters}

\begin{figure}
\begin{center}
\includegraphics[width=8cm]{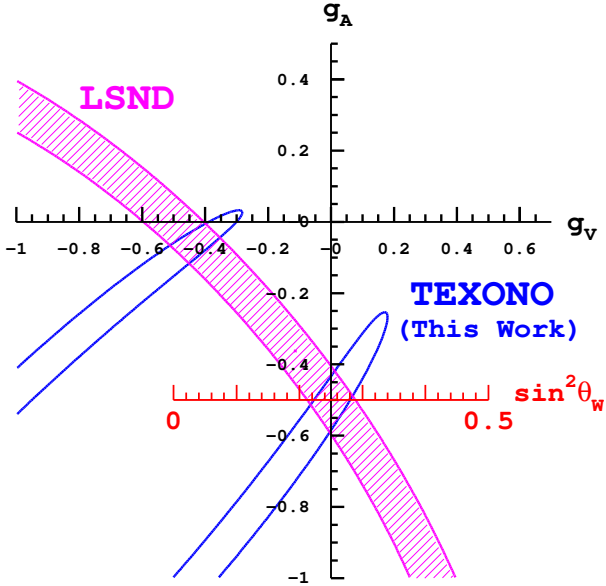}
\caption{\label{fig_gvga}
Best-fit results 
in $( g_{V} , g_{A} )$ space and in the $\s2tw$ axis
from this experiment on $\nuebar -$e and 
the LSND experiment on $\nue -$e.
The allowed regions are defined by
their corresponding statistical uncertainties.
}
\end{center}
\end{figure}

The constraints on the coupling constants $(g_V , g_A)$
were derived 
by a  minimum-$\chi ^2$ two-parameter 
fit on Eq.~\ref{eq::gvga},
with a formulation similar to that 
of Eq.~\ref{eq::chi2}.
The allowed regions 
are presented in Figure~\ref{fig_gvga}.
Results from the accelerator experiment with 
$\nue$~\cite{lsnd} are overlaid.
The complementarity of the $\nue -$e and $\nuebar -$e 
processes is readily seen.

The weak mixing angle was derived with
best-fit on Eq.~\ref{eq::s2tw}, giving
\begin{equation}
{\rm
\s2tw=0.251 \pm 0.031 ({\it stat}) \pm 0.024 ({\it sys})  
}
\end{equation}
at $\chi ^2$/dof=8.7/9, 
in excellent agreement with the SM value of 
$\s2tw (SM) = 0.23867 \pm 0.00016$ 
at this low $Q^2$ (${\rm < 10^{-4} GeV ^2}$) range~\cite{smq2}.
The combined 
uncertainty of $\pm$0.039 from this 
measurement is less than
that from the LSND accelerator $\nue -$e experiment
of $\pm$0.051.
The improvement is due to the enhancement factors
favoring $\nuebar -$e processes, as indicated in
Eq.~\ref{eq::ds2tw}.

\begin{figure}
\begin{center}
\includegraphics[width=8cm]{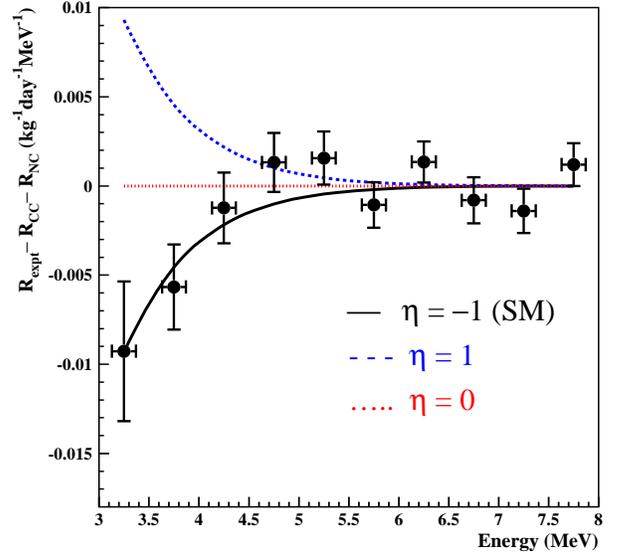}
\caption{\label{fig_inter}
The measurement of interference term from best-fit
to the data in the 3$-$8 MeV energy range.
The solid line corresponds to
the SM expectation of $\eta = - 1$.
}
\end{center}
\end{figure}

The interference term 
was probed using Eq.~\ref{eq::interf}.
The best-fit value of the sign-parameter $\eta$ is 
\begin{equation}
\eta = - 0.92 \pm 0.30 (stat) \pm 0.24 (sys) 
\end{equation}
at $\chi ^2$/dof=8.8/9.
The residual spectrum showing 
$( R_{expt} - R_{\rm CC} - R_{\rm NC} )$ 
is displayed in Figure~\ref{fig_inter},
with the expected spectra for $\eta = 0, \pm 1$ 
overlaid.
The results verified destructive interference
in the SM $\nuebar -$e interactions.

\subsection{Neutrino Electromagnetic Properties}

To include possible effects due to $\munu$ and $\nuchrad$,
the expression of Eq.~\ref{eq_sys} was modified to
\begin{equation}
{\it R} ( \munu ; \nuchrad) =
R_{\rm H1} (\rm ON) - 
[ {\it R}_{SM} ( \nu ) + {\it R}_{H1} (BKG) ]
~~ .
\label{eq_rate_em}
\end{equation}

The non-standard interaction parameter $\nuchrad$ 
as defined in Eq.~\ref{eq::qrad}
was measured to be
\begin{equation}
\nuchrad = [ 0.61 \pm 1.30 (stat) \pm 1.01 (sys) ] 
\times 10^{-32} ~{\rm cm^{2}} 
\label{eq_charge_rad2}
\end{equation}
at $\chi^{2}$/dof=8.7/9. This can be translated to bounds of 
\begin{equation}
-2.1 \times 10^{-32} ~{\rm cm^{2}} 
~ < ~ \nuchrad ~ <  ~
3.3 \times 10^{-32} ~{\rm cm^{2}} 
\end{equation}
at 90\% confidence level, an improvement over the
current limits by the LSND experiment with $\nue -$e~\cite{lsnd}:
$-2.97 \times 10^{-32} ~{\rm cm^{2}} 
< \nuchrad < 
4.14 \times 10^{-32} ~{\rm cm^{2}} $.

Similarly, the best-fit value for $\munu ^ 2$ is
\begin{equation}
\munu ^{2} = [ 0.42 \pm 1.79 (stat) \pm 1.49 (sys) ] 
\times 10^{-20} ~ \mu_{\rm B} ^2
\end{equation}
at $\chi^{2}$/dof=8.7/9, 
which corresponds to a limit of
the $\nuebar$ neutrino magnetic moment of 
\begin{equation}
\mu_{\bar{\nu}_{e}} < 2.2 \times 10^{-10} ~ \mu_{\rm B}
\end{equation}
at 90\% confidence level.
This is, however, less stringent than the
best published limit of 
$\mu_{\bar{\nu}_{e}} < 0.74 \times 10^{-10} ~ \mu_{\rm B}$
with germanium detector at 12~keV 
analysis threshold~\cite{texonomunu}.

\section{Summary and Prospects}

We report in this article an improved measurement
of reactor $\nuebar$ with the atomic electrons
at the $\rm{Q^2} \sim 10^{-6} ~ GeV^2$ range.
Complementary and comparable sensitivities
on the SM electroweak parameters were achieved 
as those measurements with accelerator $\nue$ at
higher $\rm{Q^2}$.

The detector concept allowed complete three-dimensional
event reconstruction, with which we
demonstrated that the background above 3~MeV
could be identified, studied and accounted for to the
level of $\sim$1\% accuracy. The background understanding
and subtraction enhanced the experimental 
sensitivities beyond the conventional Reactor ON$-$OFF
comparisons.

\begin{table}
\caption{
Projected statistical sensitivities
on $\xi$ and $\s2tw$
under various realistically
achievable improvement to
the experiment.
}
\label{tab_prosensit}
\begin{ruledtabular}
\begin{tabular}{lcc}
Improvement& $\Delta_{stat} ( \xi )$  & $\Delta_{stat} [ \s2tw ]$  \\ \hline
This Work &  0.21 & 0.031 \\
\multicolumn{3}{l}{\underline{Improved Feature} :} \\
~ A. $\times 10$ Data Strength & 0.07 & 0.010 \\
~ B. Background Reduction & & \\
~~~ B1: $>$99\% Cosmic-Ray Efficiency & 0.12 & 0.018 \\
~~~ B2: $\times \frac{1}{10}$ Reduction in &  &  \\
~~~ \hspace*{0.5cm} Ambient \& $^{208}$Tl $\gamma$'s & 0.16 & 0.024 \\
~ $\ast$ With Both B1+B2 & 0.05 & 0.007 \\
All Features A+B1+B2 Combined  & 0.015 & 0.0022
\end{tabular}
\end{ruledtabular}
\end{table}

The sensitivities can be further enhanced. As illustrations,
the projected improvement under various realistically
achievable assumptions are summarized in 
Table~\ref{tab_prosensit}.
Electromagnetic calorimeters using CsI(Tl) with 
tens of tons of mass have been constructed, such that
the target mass is easily expandable.
As shown in Figure~\ref{bkgchannel},
the dominant background above 3~MeV were all external
to the target scintillator. 
Accordingly, they will be attenuated 
effectively through self-shielding in a target with
bigger mass.
The incorporated features 
listed in Table~\ref{tab_prosensit} 
correspond to 10 times increase
in data strength
and $>$10~times suppression in background.
The statistical accuracies 
can be improved to 
1.5\% and 0.9\%
for $\xi$ and $\s2tw$, respectively.

As indicated in Table~\ref{syserr},
the systematic errors on background subtraction 
are related to the actual background level, such that
they will also contribute to $\Delta_{sys} ( \xi )$ at
the $\alt 2\%$ level 
under the assumption of Table~\ref{tab_prosensit}. 
Modest improvement on
the evaluation of reactor neutrino spectra 
will attain similar accuracy.
To eliminate the errors in fiducial mass,
active light guides with a different scintillating crystal 
can be coupled to both ends of the CsI(Tl) target crystal.
A good candidate is the pure CsI crystal.
The vast difference in the scintillation decay times 
($\sim$10~ns versus $\sim$1000~ns 
for CsI and CsI(Tl), respectively)~\cite{csiproto}
makes the definition of an inner target volume simple
and exact using PSD techniques.

The projected sensitivities of
such experiments are similar to 
those estimated with a large liquid scintillator 
target~\cite{nueliqscin}, and
can complement the $\s2tw$ measurements with the 
high energy accelerator experiments.
In particular, these experiments can probe
the anomalous NuTeV results~\cite{nutev} 
at comparable
sensitivities but with a different neutrino 
interaction channel
and at a low $\rm{Q^2}$~\cite{rosner04}.

\section{Acknowledgments}

The authors are indebted to the many colleagues who
made this experiment possible. The
invaluable contributions by the technical staff
of our institutes and of the Kuo-Sheng Nuclear Power 
Station are gratefully acknowledged. The
veto scintillators loan from the CYGNUS
Collaboration is much appreciated.
We appreciate comments from Prof. R. Shrock 
on neutrino charge radius.
This work is supported by fundings provided 
by the National Science Council and the Academia Sinica,
Taiwan under various contracts, 
the National Science Foundation, China,
under contract 19975050, 
as well as TUBITAK,
Turkey, under contract 108T502.

\end{document}